%% file: main.tex
\documentclass[acmsmall]{acmart}

\setcopyright{rightsretained}

\usepackage{hyperref} 

\usepackage{amsmath}
\usepackage{amstext}
\usepackage{amsthm}
\usepackage{array}
\usepackage{booktabs}
\usepackage{caption}
\usepackage{cleveref}
\usepackage [autostyle, english = american]{csquotes}
\usepackage{enumerate}
\usepackage{geometry}
\usepackage{graphicx}
\usepackage[latin1]{inputenc}
\usepackage{listings}
\usepackage{mathpartir}
\usepackage{mdframed}
\usepackage{moreverb}
\usepackage{relsize}
\usepackage{subcaption}
\usepackage{stmaryrd}
\usepackage{tikz}
\usepackage{url}
\usepackage{xcolor}
\usepackage{xparse}
\usepackage{xspace}

\usetikzlibrary{positioning}
\usetikzlibrary{shapes.misc}
\usetikzlibrary{calc}

\colorlet{listingSpecial}{violet!70!black}

\NewDocumentEnvironment{mathfigure}{mm}
{\figure 
\mdframed
\centering
\csname gather*\endcsname  
}
{
\csname endgather*\endcsname
\caption{#2}
\label{#1}
\endmdframed
\endfigure}

\MakeOuterQuote{"}

\newtheorem{theorem}{Theorem}

\newcommand{\arity}{\mathsf{N}}
\newcommand{\body}{\mathsf{body}}
\newcommand{\BNF}{\quad\operatorname{::=}\quad}
\newcommand{\BNFOR}{\quad\operatorname{\big{|}}\quad}
\newcommand{\BSemantics}[2]{\left\llbracket #1 \right\rrbracket_{#2}}

\DeclareMathOperator*{\conjunction}{\mathlarger{\mathlarger{\mathlarger{\land}}}}
\newcommand{\Dist}{\mathrm{Dist}}
\newcommand{\effects}{\mathbb{S}}
\newcommand{\eg}{\textit{e.g.}\xspace}

\newcommand{\LangSys}{\textsf{AsInst}\xspace}
\newcommand{\LangOp}[1]{\textcolor{teal!65!black}{\ensuremath{\mathsf{#1}}}}
\newcommand{\LangTrColor}[1]{\textcolor{red!70!black}{\ensuremath{#1}}}
\newcommand{\LangTr}{\LangTrColor{\mathsf{tr}}}
\newcommand{\LangGuard}[1]{{#1}}
\newcommand{\identifier}{\mathsf{id}}
\newcommand{\ie}{\textit{i.e.}\xspace}
\newcommand{\ithen}{\mathbin{\raisebox{0.2ex}{\scalebox{1.0}[0.8]{$\mathnormal|$}}}}

\newcommand{\maximum}{\mathrm{max}}
\newcommand{\nb}{\textit{n.b.}\xspace}

\newcommand{\Probability}[1]{\mathcal{P}\!\mathit{r}\!\bigl[\, #1 \,\bigr]}
\DeclareMathOperator*{\product}{\mathlarger{\mathlarger{\mathlarger{\times}}}}

\newcommand{\singleton}{\mathrm{singletonDist}}
\DeclareMathOperator*{\sumation}{\mathlarger{\mathlarger{\mathlarger{+}}}}
\newcommand{\Support}{\mathcal{S}\!\mathit{upp}}
\newcommand{\tims}{\mathsf{time}}

\newcommand{\Timestamps}{\mathit{Time}}
\newcommand{\tplus}{\mathbin{\widehat{+}}}
\newcommand{\truncate}[2]{\left\lceil #1 \right\rceil_{#2}}
\newcommand{\truncateB}[2]{\left\lceil\!\!\left\lceil #1 \right\rceil\!\!\right\rceil_{#2}}
\newcommand{\Types}{\mathbb{T}}
\newcommand{\typOUT}{\mathsf{typOUT}}
\newcommand{\typsIN}{\mathsf{typsIN}}
\newcommand{\tzero}{\mathtt{z \mkern -2.1mu e \mkern -2.0mu r \mkern -1.8mu o}}
\newcommand{\varsIN}{\mathsf{varsIN}}
\newcommand{\WithTimestamp}[1]{\widehat{#1}}
\newcommand{\WithTimestamps}[1]{\left\langle #1 \right\rangle}

\begin{document}

\title{Probabilistic, Resource-Aware, Asynchronous, Out-of-Order Choreographies}

\author{Mako Bates}
\affiliation{%
  \institution{University of Vermont}
  \city{Burlington}
  \country{USA}
}
\email{Mako.Bates@uvm.edu}

\author{Steven Baldasty}
\affiliation{%
  \institution{University of Vermont}
  \city{Burlington}
  \country{USA}
}
\email{Steven.Baldasty@uvm.edu}

\author{Ernest Hyun}
\affiliation{%
  \institution{University of Vermont}
  \city{Burlington}
  \country{USA}
}
\email{Ern.Kim@uvm.edu}

\author{Christian Skalka}
\affiliation{%
  \institution{University of Vermont}
  \city{Burlington}
  \country{USA}
}
\email{Christian.Skalka@uvm.edu}

\author{Joseph P. Near}
\affiliation{%
  \institution{University of Vermont}
  \city{Burlington}
  \country{USA}
}
\email{jnear@uvm.edu}

\begin{abstract}
Futures-based implementations of out-of-order choreographies can substantially improve latency and throughput, but their actual behavior depends on resources such as communication delay, computation time, failures, and recovery. Existing formal models such as Ozone's $O_3$ describe which executions are possible, but do not directly explain how likely those executions are or how long they take. In this work we present \LangSys, a probabilistic, resource-aware language for modeling the semantics of asynchronous choreographies with out-of-order execution. \LangSys programs are interpreted as temporal Bayesian networks that model both the values produced at runtime and the times at which they become available. We prove that this central semantics correctly captures a corresponding futures-style network semantics. We also show that \LangSys can encode Ozone-style select-and-merge conditionals, and we use case studies to model communication-failure recovery and analyze runtime performance.
\end{abstract}



\maketitle

\section{Introduction}
\label{sec:introduction}

Out-of-order execution is an important source of performance in distributed systems. When communication and computation can overlap, systems can adapt to the order in which messages arrive instead of waiting for a fixed global schedule. This is especially appealing in choreographic programming, where one would like to retain a high-level account of a distributed protocol while still exploiting the asynchronous behavior of the underlying network and runtime.

Ozone~\cite{ozone} introduced $O_3$, a formal model of fully out-of-order choreographies, and demonstrated a futures-based implementation technique that can substantially reduce latency and improve throughput. But the formal semantics of $O_3$ is nondeterministic: it describes which executions are possible, not how likely they are or how long they take. As a result, there remains a gap between the formal model and the actual futures-based implementation when the goal is to reason about resource usage and runtime behavior.

In this work we present \LangSys, a resource-aware probabilistic language for modeling the semantics of asynchronous choreographies with out-of-order execution. \LangSys aims to model the actual futures-based execution of systems like Ozone, and to enable reasoning about the performance of these systems. \LangSys programs are written as asynchronous instructions whose effects produce both values and availability times. In \LangSys, time is the primary resource: effect definitions can model network latency, computation delay, omission failures, timeouts, and recovery behavior. The central semantics interprets programs as temporal Bayesian networks, and the network semantics describes futures-based distributed execution in terms of asynchronously evolving world states. Together these semantics let us reason about the performance of asynchronous choreographies in a way that matches futures-style execution more closely than a purely nondeterministic semantics.

\LangSys is also intended as a small target language for modeling other choreographic systems. We give a translation from an Ozone-style language with select-and-merge conditionals, showing how \LangSys can model knowledge of choice using ordinary data and timing dependencies. We then use case studies and benchmarks to show how its probabilistic semantics can model both communication-failure recovery and the performance of the resulting systems. We present empirical results demonstrating how the \LangSys semantics can produce concrete performance analysis results.

\paragraph{Contributions.}
This paper makes the following contributions:
\begin{itemize}
\item We introduce \LangSys the first resource-aware probabilistic language for modeling asynchronous choreographies (Section~\ref{sec:by-example}).
\item We present a single-forward-pass Bayes-net semantics for \LangSys that models both runtime values and the times at which they become available, and show that it agrees with a distributed futures-based time-step semantics (Appendix~\ref{sec:proofs}).
\item We give a translation from an Ozone-style language with select-and-merge knowledge of choice, demonstrating that \LangSys can encode structured conditionals using guarded asynchronous instructions (Section~\ref{sec:embedding}).
\item We present case studies showing how \LangSys can model systems that recover from communication failures (Section~\ref{sec:beyond_koc}).
\item We give empirical results demonstrating how the semantics can be used to model the performance of the case-study systems (Section~\ref{sec:benchmark-runtime}).
\end{itemize}

\section{\LangSys by Example}
\label{sec:by-example}

\LangSys is designed as a minimal model for out-of-order asynchronous choreographies. In \LangSys, each variable has both a \emph{value}
and a \emph{timestamp}
(written $x = (v,t)$ or $\LangOp{val}(x) = v, \LangOp{time}(x) = t$),
and a choreography is written as a sequence of instructions.
Each instruction invokes an externally defined \emph{effect} (e.g. the \textsf{com} effect for communication).
Each effect specifies a joint probability distribution over returned values and completion times.
This makes it possible to describe communication, local computation, and futures-style waiting within a single language.

The key semantic idea is that an \LangSys program denotes not just a set of possible executions (as in $O_3$'s nondeterministic semantics), but a probability distribution over executions. Explicit tracking of timestamps allows the semantics to directly answer questions about how quickly a choreography runs. Our semantics also allows probabilistic behavior in other aspects of execution, such as values themselves.

\subsection{Example: Producer/Consumer}

As a running example, in Figure~\ref{fig:ozone-asinst-example} we reproduce the producer/consumer choreography from Figure~1 of Plyukhin et al.~\cite{ozone}. In this choreography, two producers $p_1$ and $p_2$ send values to a shared consumer $q$. The consumer computes on each received value and sends the results back. The example motivates out-of-order execution: a sequential choreography requires $q$ to process $p_1$'s message first, even if it arrives later than $p_2$'s; an out-of-order choreography allows $q$ to process the messages in either order (or even in parallel).

\LangSys represents this behavior as a list of instructions. Each instruction applies an \emph{effect} and binds the future produced by that effect.
The probability distribution of an effect's outcome depends on (is a function of) both the value and time components of its inputs.
These distributions may represent network latency, may encapsulate complicated re-try behavior or assumptions about long-running computations,
or they may be point (singleton) distributions if a process is deterministic.

\begin{figure}
\centering
\begin{minipage}[t]{0.47\textwidth}
\textbf{Ozone choreography}
\begin{lstlisting}
1 : p1.produce() -> val q.x1;

2 : p2.produce() -> val q.x2;

3 : q.compute(q.x1) -> val p1.y1;

4 : q.compute(q.x2) -> val p2.y2
\end{lstlisting}
\end{minipage}
\hfill
\begin{minipage}[t]{0.47\textwidth}
\textbf{\LangSys program}
\begin{lstlisting}
x1p <- produce();
x1  <- com[p1,q](x1p);
x2p <- produce();
x2  <- com[p2,q](x2p);
y1q <- compute(x1);
y1  <- com[q,p1](y1q);
y2q <- compute(x2);
y2  <- com[q,p2](y2q)
\end{lstlisting}
\end{minipage}
	\caption{The Ozone producer/consumer choreography from Figure~1 of Ozone~\cite{ozone}, and the corresponding \LangSys program (with trivial instruction guards omitted).}
\Description{Two side-by-side code listings. The left listing shows the four-line Ozone producer consumer choreography. The right listing shows the corresponding AsInst program with produce, compute, and communication effects.}
\label{fig:ozone-asinst-example}
\end{figure}

\Cref{fig:ozone-asinst-example} expands each Ozone communication into explicit \LangSys effects. For example, line~1 of the Ozone program says that $p_1$ produces a value and sends it to $q$. In \LangSys this becomes a production effect, followed by a communication effect whose static parameters identify the sender $p_1$ and receiver $q$.

\subsection{Defining Effects}

Effects in \LangSys are probabilistic functions that define both value and time aspects of their results.
An effect argument is either a timestamped value $(v,t)$, where $v$ is the value and $t$ is the timestamp, or $\Diamond$, which indicates that the argument is never available (e.g. due to omission failure).
In addition to its probabilistic arguments, an effect may have some static parameters written in square brackets;
these may be read as indexing into a (possibly infinite) family of effects.
The static arguments are often used to list the participants or other fixed data that selects a particular member of the effect family.
When defining effects, we write $\Dist(\tau)$ for the set of probability distributions over $\tau$. A unary effect is a probabilistic function of the form $f : (\tau_1 \times \Timestamps) \cup \Diamond \to \Dist((\tau_2 \times \Timestamps) \cup \Diamond)$, which maps timestamped inputs to distributions over timestamped outputs or $\Diamond$. In definitions, $x \leftarrow \mu$ denotes sampling $x$ from distribution $\mu$, and $\mathbf{return}\; v$ denotes the point mass distribution that assigns probability 1 to the value $v$.

For example, the definition of $\LangOp{produce}$ appears below.
It produces integers whose values are drawn from some distribution $P$.
The output value has timestamp $1$, reflecting the fact that values are produced quickly and do not depend on any other event.
Our semantics requires each effect to take non-zero time; by convention, we use a timestamp of $1$ for effects like $\LangOp{produce}$ that take negligible time.
\LangSys does not assign any particular unit to timestamps, and the unit for an individual tick can be set arbitrarily small (e.g. milliseconds or nanoseconds).
%
\[
\boxed{
\begin{aligned}
&\LangOp{produce} &:&\quad () \to \Dist(\mathit{Int} \times \Timestamps) \\
&\LangOp{produce}() &=&\quad v \gets P \\[-4pt]
&&&\quad \textbf{return}\; (v, 1) \\
\end{aligned}
}
\]

The definition of $\LangOp{compute}$ appears below. Its output value is the result of calling some function $f$ on the value of the input,
and the timestamp of the output is immediately after the timestamp of the input.
We require all effects to take non-zero time; longer computation delays can be modeled deterministically or probabilistically as part of the effect's definition.
\[
\boxed{
\begin{aligned}
&\LangOp{compute} &:&\quad (\mathit{Int} \times \Timestamps) \cup \Diamond \to \Dist((\mathit{Int} \times \Timestamps) \cup \Diamond) \\
&\LangOp{compute}((x, t)) &=&\quad \textbf{return}\; (f(x), t+1) \\
&\LangOp{compute}(\Diamond) &=&\quad \textbf{return}\; \Diamond \\
\end{aligned}
}
\]
Like Ozone, \LangSys does not provide a complete language of expressions. Instead, we assume that local computation can be expressed as an effect in a straightforward way. Specifically, any side-effect-free expression can be compiled into an effect written as a function of its free variables, in the same way as $\LangOp{compute}$.

The definition of $\LangOp{com}$ appears below.
Its static arguments identify the sender $p$ and receiver $q$, and its output value is equal to its input value.
The timestamp of the output is equal to the timestamp of the input plus delay introduced by the latency on the channel between $p$ and $q$,
described by some distribution $\Lambda(p,q)$.
\[
\boxed{
\begin{aligned}
&\LangOp{com}[p,q] &:&\quad (\tau \times \Timestamps) \cup \Diamond \to \Dist((\tau \times \Timestamps) \cup \Diamond) \\
&\LangOp{com}[p,q]((x, t)) &=&\quad d \gets \Lambda(p, q) \\[-4pt]
&&&\quad \textbf{return}\; (x, t+d) \\
&\LangOp{com}[p,q](\Diamond) &=&\quad \textbf{return}\; \Diamond \\
\end{aligned}
}
\]

\subsection{Knowledge of Choice}

\begin{figure}
\centering
\begin{minipage}[t]{0.47\textwidth}
\textbf{Selective choreography}
\begin{lstlisting}
1: Buyer.title -> Seller.x;

2: Seller.getPrice(x) -> Buyer.price;

3: if Buyer.check(price) then

4:   Buyer -> Seller[ok];
5:   Buyer.address -> Seller.y; 

6:   Seller.getDate(y) -> Buyer.date
7: else
8:   Buyer -> Seller[ko]
\end{lstlisting}
\end{minipage}
\hfill
\begin{minipage}[t]{0.47\textwidth}
\textbf{\LangSys program}
\begin{lstlisting}
x <- com[Buyer,Seller](title);
priceS <- getPrice(x);
price <- com[Seller,Buyer](priceS);
// conditional
priceOk <- check(price);
gt <- guardT(priceOk);
gf <- guardF(priceOk);
// "then" branch
gt | ok <- com[Buyer,Seller]("ok");
gt | y <- com[Buyer,Seller](address);
ok | dateS <- getDate(y);
ok | date <- com[Seller,Buyer](dateS);
// "else" branch
gf | ko <- com[Buyer,Seller]("ko")
\end{lstlisting}
\end{minipage}
\caption{The classic bookseller protocol (example 6.12 from~\cite{montesi_book}), and the corresponding \LangSys program. This example demonstrates encoding select-based knowledge of choice in \LangSys.}
\Description{Two side-by-side code listings. The left listing shows a selective bookseller choreography with buyer and seller messages, a conditional, and ok or ko selections. The right listing shows the corresponding AsInst program using communication, check, guardT, guardF, and getDate effects.}
\label{fig:bookseller}
\end{figure}

\LangSys itself has no branching syntax, so it does not need a primitive projection rule that ensures knowledge of choice. However, choreographies with conditionals can be translated into \LangSys choreographies that encode the same knowledge of choice using ordinary data and timing dependencies. Section~\ref{sec:embedding} describes this approach for Ozone-style select-and-merge knowledge of choice.

As a simple example, consider the classic bookseller protocol shown in Figure~\ref{fig:bookseller}. In this choreography, the Buyer communicates knowledge of choice for the condition in line 3 to the Seller via selections on lines 4 and 8. Our encoding in \LangSys uses several new effects for local computations (\LangOp{getPrice}, \LangOp{check}, and \LangOp{getDate}), defined in the same way as $\LangOp{compute}$. The source choreography or translation discipline determines which participant performs each local computation; that participant is not a static parameter of the local computation effect.

To encode conditionals and knowledge of choice in \LangSys, we use \emph{guarded instructions} written $g \mid x \leftarrow E$, where $g$ is a variable name representing the guard. In \LangSys, a guarded instruction may execute only after the guard $g$ has arrived,
so the timestamp of the bound variable must be later than the timestamp of $g$.
Unlike traditional guarded commands, the \emph{value} of $g$ has no semantic effect;
in \LangSys, guards determine \emph{when} instructions are evaluated, via their timestamps.
They secondarily control \emph{whether} instructions are evaluated, because if $g = \Diamond$ then the instruction will never execute.

The \LangOp{guardT} and \LangOp{guardF} effects, defined below, translate boolean values into \LangSys guards. The effects use the boolean input to return either a timestamped unit value or $\Diamond$. The \LangOp{guardT} effect, for example, returns a value only when its argument is true, and returns $\Diamond$ otherwise; its result can be used as a guard for other instructions. We often bind their results to the variables $g_t$ and $g_f$.
Both outputs have the type (and therefore value) $\mathit{Unit}$,
because these variables are intended to be used as guards, so their values are not important.
Together, they translate a boolean condition from a value to timestamp-based guards which can be used to encode the branches of a conditional.
\[
\boxed{
\begin{aligned}
&\LangOp{guardT} &:&\quad (\mathit{Bool} \times \Timestamps) \cup \Diamond \to \Dist((\mathit{Unit} \times \Timestamps) \cup \Diamond) \\
&\LangOp{guardT}((\LangOp{true}, t)) &=&\quad \textbf{return}\; ((), t) \\
&\LangOp{guardT}((\LangOp{false}, t)) &=&\quad \textbf{return}\; \Diamond \\
&\LangOp{guardT}(\Diamond) &=&\quad \textbf{return}\; \Diamond
\end{aligned}
}
\]
\quad\quad
\[
\boxed{
\begin{aligned}
&\LangOp{guardF} &:&\quad (\mathit{Bool} \times \Timestamps) \cup \Diamond \to \Dist((\mathit{Unit} \times \Timestamps) \cup \Diamond) \\
&\LangOp{guardF}((\LangOp{true}, t)) &=&\quad \textbf{return}\; \Diamond \\
&\LangOp{guardF}((\LangOp{false}, t)) &=&\quad \textbf{return}\; ((), t) \\
&\LangOp{guardF}(\Diamond) &=&\quad \textbf{return}\; \Diamond
\end{aligned}
}
\]
We also use a \LangOp{first} effect to join alternatives.
It returns whichever argument becomes available first, and returns $\Diamond$ only when neither argument becomes available.
When both arguments become available at the same time, \LangOp{first} deterministically chooses the first argument.
\[
\boxed{
\begin{aligned}
&\LangOp{first} &:&\quad ((\tau \times \Timestamps) \cup \Diamond) \times ((\tau \times \Timestamps) \cup \Diamond) \to \Dist((\tau \times \Timestamps) \cup \Diamond)\\
&\LangOp{first}((x_1,t_1),(x_2,t_2)) &=&\quad \textbf{return}\; \begin{cases}
(x_1,t_1) & \text{if } t_1 \le t_2 \\
(x_2,t_2) & \text{otherwise}
\end{cases}\\
&\LangOp{first}(\Diamond,(x,t)) &=&\quad \textbf{return}\; (x,t)\\
&\LangOp{first}((x,t),\Diamond) &=&\quad \textbf{return}\; (x,t)\\
&\LangOp{first}(\Diamond,\Diamond) &=&\quad \textbf{return}\; \Diamond
\end{aligned}
}
\]
These guard effects provide variables we can use in guarded instructions
like the one on line 4 in Figure~\ref{fig:bookseller}.
Using $\mathsf{gt}$ as the guard in this instruction means that the ``ok'' selection label
will be sent from the Buyer to the Seller only if \LangOp{check} returns \LangOp{true};
otherwise, the message will never be sent.

To ensure knowledge of choice for participants who do not have access to the original condition, we can use the received selection label itself as the guard, as in line 6 of Figure~\ref{fig:bookseller}. If the condition is \LangOp{true}, then the selection will be sent, and the Seller's part of the ``then'' branch will execute after they receive the ``ok'' selection label. If the condition is \LangOp{false}, then the Seller will never receive the ``ok'' selection label, and the Seller's part of the ``then'' branch will never execute. Instead, the Seller will receive the ``ko'' selection label (line 8) and the choreography will end.

This approach effectively ``flattens'' structured conditionals, and guards the instructions in each branch with variables that encode the original control flow. We present a translation from Ozone choreographies to \LangSys that uses this approach in Section~\ref{sec:embedding}.

\subsection{Ensuring Safety}

\paragraph{Ensuring communication integrity.}
Out-of-order choreographies need a mechanism to avoid communication integrity violations (CIVs): the potential assignment of a message to the wrong variable when messages are reordered. Following Ozone, implementations may use source line numbers as \emph{integrity keys} to avoid CIVs. In our example, communication effects are written using $\LangOp{com}[p,q]$; the integrity key is not an additional static parameter in the \LangSys program. At runtime, the value of the integrity key may still be sent along with the message itself, allowing the recipient $q$ to associate the received message with the correct syntactic variable from the original choreography. Ozone's integrity keys are slightly more complicated, to handle procedures and recursion, which \LangSys lacks. \LangSys does not prescribe any particular solution for ensuring communication integrity, and other solutions are possible.

\paragraph{Ensuring accessibility.}
Accessibility is the requirement that a participant can only use data that is actually available at that participant. This discipline is not part of the core effect type syntax. Instead, it is enforced outside of \LangSys, by the source-language type system or translation, or a separate well-formedness check that tracks which participant is allowed to read each variable. For example, a local computation executed by participant $p$ should only depend on variables available to $p$, and communication effects such as $\LangOp{com}[p,q]$ are the points where the projected network may transfer data between participants. If an effect signature or translation lies about which variables a participant can access, the resulting \LangSys program may still have an \LangSys interpretation, but may not be realizable by a distributed network.

\paragraph{Handling communication failures.}
\LangSys models omission failures (and omissions of any other kind)
by extending the set of possible results of all effects with a placeholder value $\Diamond$, read as \emph{"nothing"} or \emph{"never"}.
A communication effect that returns $\Diamond$ is one whose message is never received, either due to an omission failure or because it was never sent in the first place.
Effects that may never produce a result include $\Diamond$ in their codomain, while other effects can still consume $\Diamond$ and react at finite times.
For example, a timeout effect with deadline $d$ can become available after $d$ ticks and report whether its argument arrived before the deadline,
as shown below.
Note that the possibility of $\Diamond$ values requires that most effects will be defined case-wise to handle $\Diamond$ inputs specifically.
\[
\boxed{
\begin{aligned}
&\LangOp{timeout}[p,d] &:&\quad (\tau \times \Timestamps) \cup \Diamond \to \Dist(\mathit{Bool} \times \Timestamps) \\
&\LangOp{timeout}[p,d](\Diamond) &=&\quad \textbf{return}\; (\LangOp{false}, d)\\
&\LangOp{timeout}[p,d]((x, t)) &=&\quad \textbf{return}\; \begin{cases}
	(\LangOp{true}, t + 1) & \text{ if } t < d \\
	(\LangOp{false}, d) & \text{ otherwise}
\end{cases}
\end{aligned}
}
\]
If $x$ never arrives, or arrives after $d$, the timeout returns false at time $d$ rather than waiting forever.

\paragraph{Deadlock freedom.}
Deadlock freedom by construction is an important feature of choreographic programming languages, and usually means that a well-formed choreography projects to a set of processes that cannot reach a configuration in which every participant is stuck waiting for communication.
\LangSys is not a traditional choreographic language, and is designed to be used to reason about choreographies written in higher-level languages that already guarantee deadlock-freedom.
Since \LangSys has no loops or recursion, and the data flow of any well-formed \LangSys program is a directed acyclic graph, well-formed \LangSys programs cannot get stuck.
However, \LangSys can model systems that get stuck waiting for communications that fail, using $\Diamond$.

The theorem below captures the corresponding probabilistic guarantee.
If every effect in the program resolves with probability one whenever its arguments are available,
then every variable in the program eventually becomes available with probability one.
More generally, when effects can return $\Diamond$ to model omission failures, timeouts, or intentionally inactive branches,
\LangSys makes those assumptions explicit in the semantics.
This lets us distinguish structural deadlock from probabilistic resource failures and control-flow nonavailability,
and it lets us reason about the probability that an output appears by a deadline, never appears, or is reached through a recovery path.
\begin{theorem}\label{theorem:deadlock-freedom}
	If $\effects, \Gamma_0 \vdash C$
	and for all $S \in \effects$ and all $args \in \WithTimestamps{S.\typsIN}$
	we have that $\Diamond \not\in args$ implies $\Probability{S.\body(args) = \Diamond} = 0$,
	then for all $x \in C$ we have $\Probability{\BSemantics{C}{\effects}[x] = \Diamond} = 0$.\\
	In other words, if the effects in $C$ will always resolve provided that their arguments become available,
	then eventually all variables in $C$ will become available.\\
	The proof is by induction over the acyclic dependency structure of $C$.
\end{theorem}

\input{graphSemantics_large}

\section{Embedding Ozone / $O_3$ in \LangSys}
\label{sec:ozone-embedding}
\label{sec:embedding}

In this section, we demonstrate how \LangSys can be used to model a subset of the futures-based Ozone implementation in Choral~\cite{ozone}. We show a translation from a subset of $O_3$ into \LangSys, using communication effects for both ordinary messages and selection labels. The \LangSys semantics give translated choreographies a formal semantics that model the futures-based implementation of Ozone in Choral.
Ozone and $O_3$ provide structural conditionals and adopt the standard select-and-merge approach for knowledge of choice in choreographies. For projectable Ozone choreographies, our translation encodes select-and-merge knowledge of choice in the translated \LangSys program.
Our translation omits Ozone's choreographic functions and recursion. Modeling recursion would require extending beyond the finite Bayesian networks used here.

Figure~\ref{fig:translation} first gives the syntax for the Ozone subset we translate, then gives a direct translation into \LangSys. The instruction $\textsf{val p}.x = e$ binds a fresh variable owned by the single participant $p$ to the result of the local expression $e$. The communication instruction $\textsf{p}.e \rightarrow \textsf{val q}.x$ evaluates $e$ at $p$, sends the result to $q$, and binds the received value as $x$ at $q$. The selection instruction $\textsf{p} \rightarrow \textsf{q}[L]$ sends a label from $p$ to $q$; unlike ordinary data communication, its role is to communicate control-flow knowledge to a participant that may not know which branch is active. The conditional form binds variables functionally: $\textsf{val q}.x = \textsf{if } e@\textsf{p} \textsf{ then } C_1; \textsf{q}.x_1 \textsf{ else } C_2; \textsf{q}.x_2$ evaluates a condition known by $p$, runs one branch, and binds $x$ to the branch result at $q$.

The translation tracks control-flow knowledge with a per-participant path-condition environment. Ozone-style languages merge branches on a per-participant basis: one participant may know the branch condition directly, while another learns the choice only after receiving a selection. The Ozone translation therefore tracks $\LangGuard{\Phi}$, which maps each participant to that participant's current guard.

The source syntax in \Cref{fig:translation} is a functionalized presentation of the relevant $O_3$ fragment rather than a direct copy of the surface syntax. In ordinary statement-style choreographies, both branches of a conditional may assign to the same variable in order to give that variable a conditional value. For the translation, this creates an avoidable variable-renaming problem: \LangSys programs require each instruction to bind a fresh variable, so the translation would otherwise need to distinguish the two branch-local versions of the same source variable before joining them. We avoid this bookkeeping by presenting conditionals as if-expressions with an explicit result. In the form $\textsf{val q}.x = \textsf{if } e@\textsf{p} \textsf{ then } C_1; \textsf{q}.x_1 \textsf{ else } C_2; \textsf{q}.x_2$, the branch result variables $x_1$ and $x_2$ are distinct, and the conditional binds $x$ to whichever branch result is selected. Translating from the actual $O_3$ statement syntax to this functionalized form is a straightforward preprocessing step: choose fresh branch result names, rewrite each branch to bind its result to the corresponding fresh name, and use the conditional expression to bind the original variable name.



\begin{mathfigure}{fig:translation}{Translation of $O_3$ choreographies into \LangSys programs.}
\textbf{Syntax for the Ozone subset:}\\
\begin{align*}
C &::= 0 \mid I; C\\
I &::= \textsf{val p}.x = e
    \mid \textsf{p}.e \rightarrow \textsf{val q}.x
    \mid \textsf{p} \rightarrow \textsf{q}[L]\\
  &\;\mid\; \textsf{val q}.x =
      \textsf{if } e@\textsf{p} \textsf{ then } C_1; \textsf{q}.x_1 \textsf{ else } C_2; \textsf{q}.x_2
\end{align*}\\[10pt]
\hline
\\
\textbf{Translation to \LangSys:}\\
\begin{array}{l c l}
\LangTr(\LangGuard{\Phi}, 0) & \triangleq &
\begin{array}{l}
0\\[8pt]
\end{array}
\\
\LangTr(\LangGuard{\Phi}, \textsf{val p}.x = e; C) & \triangleq &
\begin{array}{l}
\LangGuard{\Phi}(p) \mid x \leftarrow \LangTr(p, e);\\
\LangTr(\LangGuard{\Phi}, C)\\[8pt]
\end{array}
\\
\LangTr(\LangGuard{\Phi}, \textsf{p}.e \rightarrow \textsf{val q}.x; C) & \triangleq &
\begin{array}{l}
\LangGuard{\Phi}(p) \mid u \leftarrow \LangTr(p, e);\\
u \mid x \leftarrow \LangOp{com}[p, q](u);\\
\LangTr(\LangGuard{\Phi}, C)\\[8pt]
\end{array}
\\
\LangTr(\LangGuard{\Phi}, \textsf{p} \rightarrow \textsf{q}[L]; C) & \triangleq &
\begin{array}{l}
\LangGuard{\Phi}(p) \mid m \leftarrow \LangOp{com}[p, q](L);\\
\LangTr(\LangGuard{\Phi}[q \mapsto m], C)\\[8pt]
\end{array}
\\
\LangTr(\LangGuard{\Phi}, \textsf{val q}.x = \textsf{if } e@\textsf{p} \textsf{ then } C_1; \textsf{q}.x_1 \textsf{ else } C_2; \textsf{q}.x_2; C) & \triangleq &
\begin{array}{l}
\LangGuard{\Phi}(p) \mid g \leftarrow \LangTr(p, e);\\
\LangGuard{\Phi}(p) \mid g_t \leftarrow \LangOp{guardT}(g);\\
\LangGuard{\Phi}(p) \mid g_f \leftarrow \LangOp{guardF}(g);\\
\LangTr(\LangGuard{\Phi}[p \mapsto g_t], C_1);\\
\LangTr(\LangGuard{\Phi_\Diamond}[p \mapsto g_f], C_2);\\
\LangGuard{\Phi}(q) \mid x \leftarrow \LangOp{first}(x_1, x_2);\\
\LangTr(\LangGuard{\Phi}, C)\\[4pt]
\textit{where}\; \LangGuard{\Phi_\Diamond}(r) = \Diamond\; \textit{for all $r$}
\end{array}
\end{array}
\end{mathfigure}

Figure~\ref{fig:translation} describes the translation into \LangSys. The translation for $0$ handles empty choreographies. The translations for local computation and communication produce guarded \LangSys instructions, using $\LangGuard{\Phi}(p)$ because those instructions are enabled when the sender or computing participant $p$ has learned that the current control-flow path is live. The communication translation first binds the translated local expression to a fresh variable, then communicates that variable using $\LangOp{com}[p,q]$. Both translations then translate the continuation under the same per-participant guard environment. Generated variables introduced by the translation, such as $u$, $m$, $g$, $g_t$, and $g_f$, are assumed fresh.

The translation for selections is the point where Ozone's use of select-and-merge is modeled explicitly in \LangSys. The selection $\textsf{p} \rightarrow \textsf{q}[L]$ is translated as a guarded communication of the label $L$ from $p$ to $q$. The resulting message variable $m$ is then installed as $q$'s guard for the continuation, yielding $\LangGuard{\Phi}[q \mapsto m]$. Thus, only $q$'s view of the path condition changes: participants that already knew the branch keep their existing guards, while $q$ can execute its continuation only after the selection message arrives. This mirrors merging in Ozone's endpoint projection, where each participant's behavior is merged in isolation according to the knowledge of choice available to that participant.

The translation for conditionals flattens conditionals. First, the condition expression is translated into a local effect at $p$ that produces $g$, and the \LangOp{guardT} and \LangOp{guardF} effects convert $g$ into the branch guards $g_t$ and $g_f$. The then-branch is translated with $p$'s guard updated to $g_t$, while the else-branch is translated from $\LangGuard{\Phi_\Diamond}$ with $p$ updated to $g_f$. The environment $\LangGuard{\Phi_\Diamond}$ resets every participant's path guard to $\Diamond$ before re-enabling only participant $p$ with $g_f$. This prevents participants from inheriting branch knowledge merely because the translator is processing a branch: each participant must either know the condition locally or receive an explicit selection message. Selections inside the branches propagate that knowledge to other participants by updating their entries in $\LangGuard{\Phi}$. Finally, \LangOp{first} binds the result variable $x$ to whichever branch result is available before the continuation is translated under the original guard environment.

The translation uses the communication, guard, and \LangOp{first} effects introduced earlier in the paper. Participant-specific knowledge is tracked by the translation environment rather than by target effect types.

\begin{figure}
\centering
\begin{minipage}[t]{0.47\textwidth}
\textbf{Ozone choreography}
\begin{lstlisting}
Buyer.title -> val Seller.title1;
Seller.getPrice(title1) -> val Buyer.price;
val Buyer.decision = check(price);

val Buyer.date =
  if decision@Buyer then
    Buyer -> Seller[OK];
    Buyer.addr -> val Seller.addr1;
    Seller.getDate(addr1) -> val Buyer.date1;
    Buyer.date1
  else
    Buyer -> Seller[KO];
    val Buyer.dateNone = None;
    Buyer.dateNone
\end{lstlisting}
\end{minipage}
\hfill
\begin{minipage}[t]{0.47\textwidth}
\textbf{\LangSys program}
\begin{lstlisting}
title1 <- com[Buyer,Seller](title);
priceS <- getPrice(title1);
price <- com[Seller,Buyer](priceS);
decision <- check(price);

gt <- guardT(decision);
gf <- guardF(decision);
gt | ok <- com[Buyer,Seller](OK);
gt | addr1 <- com[Buyer,Seller](addr);
ok | date1 <- getDate(addr1);
ok | date2 <- com[Seller,Buyer](date1);

gf | ko <- com[Buyer,Seller](KO);
gf | dateNone <- None();
date <- first(date2,dateNone)
\end{lstlisting}
\end{minipage}
\caption{The bookseller protocol written in the Ozone style, and the corresponding \LangSys program. This example demonstrates encoding select-based knowledge of choice in \LangSys.}
\Description{Two side-by-side code listings. The left listing shows an Ozone-style bookseller choreography where the buyer sends OK or KO selections to communicate the branch choice. The right listing shows the translated AsInst program with guard variables, guarded communications, and a first effect joining branch results.}
\label{fig:ozone_bookseller_translation}
\end{figure}

Figure~\ref{fig:ozone_bookseller_translation} shows the bookseller protocol in the Ozone style. The branch condition is known only to the buyer, so the buyer sends the selection labels \texttt{OK} and \texttt{KO} to communicate knowledge of choice to the seller. The translation converts the decision into the guard variables \texttt{gt} and \texttt{gf}; buyer-side branch instructions use those guards directly, while the seller-side date computation is guarded by the received \texttt{OK} selection message. The final \LangOp{first} instruction binds \texttt{date} to whichever branch result, \texttt{date2} or \texttt{dateNone}, is available.

\section{Communication Failures and Inferred Knowledge of Choice}
\label{sec:beyond_koc}

The examples so far use timestamps to model ordinary asynchronous execution and to encode control-flow choices. In this section, we show how \LangSys can encode protocols that are designed to recover from communication failures, which often requires \emph{inferring} knowledge of choice rather than communicating it explicitly.

\subsection{Bounded Retry}

Figure~\ref{fig:bounded-retry} shows a bounded retry protocol in which Alice sends a message to Bob, Bob sends an acknowledgement back, and Alice gives up after three total send attempts if no acknowledgement arrives in time.

\begin{figure}
\centering
\begin{lstlisting}
// First attempt: Alice sends msg and waits to d1
m1 <- com[1,Alice,Bob](msg);
a1B <- ack[Bob](m1);
a1 <- com[2,Bob,Alice](a1B);
seen1 <- timeout[Alice,d1](a1);
got1 <- guardT[Alice](seen1);
retry2 <- guardF[Alice](seen1);

// If no ack arrives by d1, send a second attempt
retry2 | m2 <- com[3,Alice,Bob](msg);
a2B <- ack[Bob](m2);
a2 <- com[4,Bob,Alice](a2B);

// Join the first two acks and wait until d2
a12 <- first[Alice](a1,a2);
seen2 <- timeout[Alice,d2](a12);
got2 <- guardT[Alice](seen2);
retry3 <- guardF[Alice](seen2);

// If still no ack arrives by d2, send a third attempt
retry3 | m3 <- com[5,Alice,Bob](msg);
a3B <- ack[Bob](m3);
a3 <- com[6,Bob,Alice](a3B);

// Alice records whether any ack arrives by d3
ack <- first[Alice](a12,a3);
success <- timeout[Alice,d3](ack)
\end{lstlisting}
\caption{A bounded retry protocol written directly in \LangSys. Alice sends at most three messages to Bob and stops retrying once an acknowledgement arrives before the next deadline.}
\Description{A code listing for a bounded retry protocol. It shows three possible Alice-to-Bob message attempts, Bob acknowledgement effects, timeout checks at deadlines d1, d2, and d3, guard variables for retry decisions, and first effects that join acknowledgements from different attempts.}
\label{fig:bounded-retry}
\end{figure}

The communication effects in this example are the same unreliable communication effects described earlier: a message may arrive after some probabilistic delay, or may be omitted by producing $\Diamond$. The local effect $\LangOp{ack}[\mathit{Bob}]$ is deterministic, like $\LangOp{compute}$: once Bob receives Alice's message, it produces an acknowledgement at Bob after a finite local delay. If Alice's message is omitted, then the input to $\LangOp{ack}$ is $\Diamond$, so the acknowledgement also remains unavailable forever.

The retry policy is expressed using ordinary \LangSys data dependencies. After the first acknowledgement variable $a1$ is created, $\LangOp{timeout}[\mathit{Alice},d1]$ produces a finite boolean at Alice: true if the acknowledgement arrived by deadline $d1$, and false otherwise. The guard variables produced by $\LangOp{guardT}$ and $\LangOp{guardF}$ turn that boolean into timing information. If Alice has already seen an acknowledgement, then $\mathit{got1}$ is available and $\mathit{retry2}$ is $\Diamond$; if not, $\mathit{retry2}$ becomes available at the timeout, enabling the second send. The third attempt is enabled in the same way after deadline $d2$. The $\LangOp{first}$ effects join acknowledgements from different attempts, so a late acknowledgement from the first attempt can still prevent the third attempt if it arrives before $d2$.

This example illustrates a case that is awkward for prior choreographic systems. A retry decision depends on negative, time-sensitive information: Alice sends again because she has not observed an acknowledgement by a deadline, not because Bob explicitly sends a failure message. Late acknowledgements also race with later retries, so the relevant behavior is not just which messages can be delivered, but when each message is delivered. Graversen et al.~\cite{graversen2017promising} encode this protocol by making communication frames explicit in the language and describing recovery for frames separately from the main choreography, for example. In \LangSys, time-aware effects enable an entirely choreographic description of the recovery strategy.


\subsection{Two-Phase Commit}

Two-phase commit is another protocol where failures affect control flow. A coordinator $C$ first asks each participant to vote on a transaction. If every participant votes yes before the vote deadline, then $C$ commits; otherwise, $C$ aborts. Figure~\ref{fig:two-phase-commit} shows a two-participant version written directly in \LangSys.

\begin{figure}
\centering
\begin{lstlisting}
// Phase 1: C asks both participants to vote
p1Prep <- com[1,C,P1](tx);
p2Prep <- com[2,C,P2](tx);
p1VoteP <- vote[P1](p1Prep);
p2VoteP <- vote[P2](p2Prep);
p1Vote <- com[3,P1,C](p1VoteP);
p2Vote <- com[4,P2,C](p2VoteP);

// C commits only if both yes votes arrive by dv
p1Yes <- yesBy[C,dv](p1Vote);
p2Yes <- yesBy[C,dv](p2Vote);
decision <- decide[C](p1Yes,p2Yes);

// Phase 2: C sends the final decision once
p1Dec <- com[5,C,P1](decision);
p2Dec <- com[6,C,P2](decision);
p1Out <- applyDecision[P1](p1Dec);
p2Out <- applyDecision[P2](p2Dec);

// Each participant acknowledges the decision once
p1AckP <- ack[P1](p1Out);
p2AckP <- ack[P2](p2Out);
p1Ack <- com[7,P1,C](p1AckP);
p2Ack <- com[8,P2,C](p2AckP);

// C records whether both acks arrive by da
p1Done <- timeout[C,da](p1Ack);
p2Done <- timeout[C,da](p2Ack);
complete <- both[C](p1Done,p2Done)
\end{lstlisting}
\caption{A two-phase commit protocol written directly in \LangSys. The coordinator commits only if both participants vote yes by the vote deadline, then sends the final decision once to each participant.}
\Description{A code listing for a two-participant two-phase commit protocol. It shows prepare messages from coordinator C to participants P1 and P2, participant votes, coordinator deadline checks and decision computation, final decision messages, participant acknowledgements, and final timeout checks for completion.}
\label{fig:two-phase-commit}
\end{figure}

The local effects in this example are deterministic computations like the effects used above. The effect $\LangOp{vote}[P_i]$ models participant $P_i$'s local vote after receiving the prepare message. The effect $\LangOp{yesBy}[C,dv]$ converts each vote into a boolean at $C$: it is true exactly when a yes vote arrives by deadline $dv$, and false if the vote is no, late, or omitted. The effect $\LangOp{decide}[C]$ returns commit only when both of these booleans are true. The effects $\LangOp{applyDecision}$ and $\LangOp{ack}$ model the participant applying the final decision and acknowledging it, and $\LangOp{both}[C]$ records whether both acknowledgements arrived by the acknowledgement deadline.

The example is intentionally simplified for brevity (e.g. by including only two participants and no retry strategy). The same structure can be combined with the bounded-retry pattern from the previous subsection for prepare, vote, decision, or acknowledgement messages. It can also be extended mechanically to any fixed number of participants by adding one vote and acknowledgement block per participant and replacing $\LangOp{both}$ with an all-participants conjunction.

Like bounded retry, two-phase commit is difficult to describe in prior choreographic systems because the interesting choices depend on the absence of messages by a deadline. The coordinator aborts not only when it receives a no vote, but also when it fails to receive a yes vote in time. Likewise, completion depends on whether acknowledgement messages arrive before the final deadline. This is the same failure-recovery setting studied by Graversen et al.~\cite{graversen2017promising}, but expressed here without adding a separate recovery language.

\section{Evaluation: Reasoning About Benchmark Runtime}
\label{sec:benchmark-runtime}

We implemented a proof-of-concept library in Python for modeling \LangSys programs. Our implementation uses the pgmpy library for causal inference to build the Bayes net representation of the program and then perform probabilistic inference to query properties of interest.
The case studies in this section use this implementation to query completion-time behavior for protocols presented earlier, including with failures, deadlines, and out-of-order execution.

\subsection{Bounded Retry}

\begin{figure}
\centering
\includegraphics[width=0.5\linewidth]{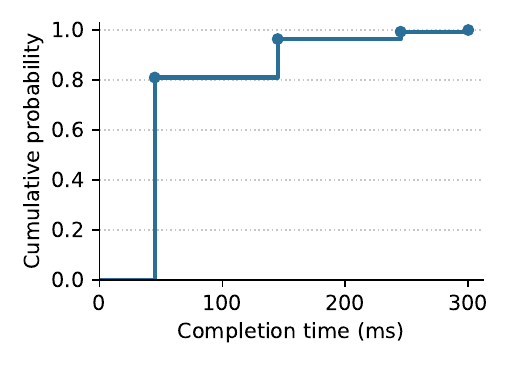}
\caption{CDF of bounded-retry completion time under a lossy channel (latency of 20ms, but 10\% packet loss). The final step corresponds to Alice reaching the last timeout.}
\Description{A step-shaped cumulative distribution plot for bounded retry completion time. The curve rises at the first, second, and third acknowledgement opportunities and reaches the remaining probability mass at the final timeout.}
\label{fig:bounded-retry-cdf}
\end{figure}

We model the bounded-retry protocol from Section~\ref{sec:beyond_koc} on a lossy channel where each message takes $20\mathrm{ms}$ to arrive; messages arrive with probability $0.9$ and are omitted with probability $0.1$. We use the \LangSys semantics to reason about how often Alice succeeds on each attempt, how often she reaches the final timeout, and how the retry policy changes the distribution of completion times. The resulting completion-time CDF is shown in \Cref{fig:bounded-retry-cdf}.

The CDF has one step for each possible attempt. Alice observes success after the first round trip at $45\mathrm{ms}$ with probability $0.81$, after the second attempt by $145\mathrm{ms}$ with cumulative probability $0.96$, and after the third attempt by $245\mathrm{ms}$ with cumulative probability $0.99$; the remaining probability mass reaches the final timeout at $300\mathrm{ms}$. This shows that the protocol usually completes quickly even though it has a recovery path, and it shows that \LangSys can quantify behavior for protocols that rely on inferred knowledge of choice in the presence of omission failures.

\subsection{Two-Phase Commit}

\begin{figure}
\centering
\begin{subfigure}[t]{0.48\linewidth}
\centering
\includegraphics[width=\linewidth]{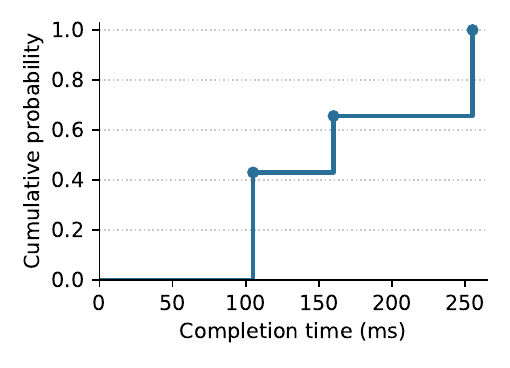}
\caption{Lossy channel}
\label{fig:two-phase-commit-cdf}
\end{subfigure}
\hfill
\begin{subfigure}[t]{0.48\linewidth}
\centering
\includegraphics[width=\linewidth]{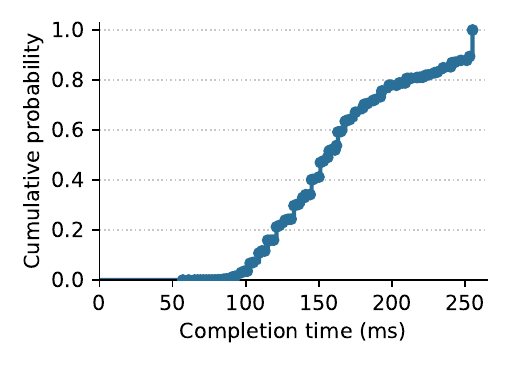}
\caption{TCP-like reliable channel}
\label{fig:two-phase-commit-tcp-cdf}
\end{subfigure}
\caption{CDFs of two-phase commit completion time under two communication models. In the lossy channel model, messages take $20\mathrm{ms}$ or are dropped; in the TCP-like model, messages are reliable with a discretized long-tailed latency distribution.}
\Description{Two side-by-side cumulative distribution plots for two-phase commit completion time. The lossy-channel plot has discrete steps caused by fixed message delay and deadline observations. The TCP-like plot has a smoother, more spread-out curve caused by reliable long-tailed latency.}
\label{fig:two-phase-commit-cdfs}
\end{figure}

We model the two-participant two-phase commit protocol from Section~\ref{sec:beyond_koc}. We evaluate the same protocol under two communication models: the lossy $20\mathrm{ms}$ channel used above, and a reliable TCP-like latency distribution with most probability mass near low latency and a longer tail. The goal is to learn how timeout-based aborts and acknowledgement deadlines affect completion time under different network assumptions. The resulting CDFs are shown in \Cref{fig:two-phase-commit-cdfs}.

Under the lossy model, the CDF has a few discrete steps because every delivered message has the same delay and omissions are observed only at deadlines. The model commits with probability $0.65$, while missing prepare or vote messages force an abort; completion-time mass at the final deadline corresponds to executions whose outcome depends on timeout observations. Under the TCP-like model, there are no omissions, but the long-tailed latency distribution spreads completion over many times. This comparison illustrates that \LangSys separates the protocol structure from the resource model: the same choreography can be reanalyzed by swapping the effect distributions.

\subsection{Concurrent Producers}

\begin{figure}
\centering
\includegraphics[width=0.5\linewidth]{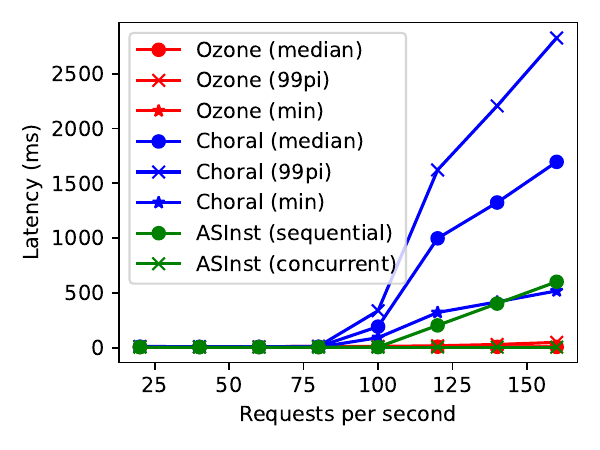}
\caption{Predictions of the AsInst implementation overlaid on the Ozone results for
\textit{Concurrent producers latency} from Figure 19(a), in addition to the \textit{minimum} latency.}
\Description{A latency plot comparing AsInst predictions with published Ozone benchmark results for concurrent producers. The AsInst prediction follows the low-latency trend and is shown alongside the minimum-latency and measured Ozone curves.}
\label{fig:ozone-result-overlay}
\end{figure}

We also model the concurrent producer/consumer benchmark from Ozone~\cite{ozone}, discussed in Section~\ref{sec:by-example}. The model asks whether the \LangSys interpretation captures the latency improvement of the out-of-order Ozone version relative to the sequential Choral version. \Cref{fig:ozone-result-overlay} overlays the \LangSys predictions with the published Ozone benchmark results.

The \LangSys prediction for concurrent execution stays close to the lowest-latency Ozone measurements, while the sequential prediction tracks the growth trend of the sequential Choral implementation. It does not match the median or tail latencies of the concrete system, because this model includes only message latency and added computation time, and omits other sources of runtime overhead. The result shows that \LangSys can predict the idealized latency induced by the choreography and effect model.




\section{Related Work}
\label{sec:related-work}

\paragraph{Choreographic programming.}
Choreographic programming represents distributed behavior as a single global program and compiles it to endpoint implementations, with classic guarantees such as communication safety and deadlock freedom by construction~\cite{montesi_book,carbone2013deadlock,cruzfilipe2023formal}. This work builds on that global-programming perspective, but it uses \LangSys as a small semantic target rather than as a source language with structured control flow. Recent choreographic languages and embeddings have expanded the programming model in several directions: Choral brings choreographies to realistic object-oriented programming~\cite{giallorenzo2024choral}; Pirouette studies higher-order functional choreographies~\cite{hirsch2022pirouette}; library-level approaches such as HasChor, ChoRus, and related work improve portability and support richer patterns such as census polymorphism~\cite{shen2023haschor, bates2025efficient}; process and location polymorphism make choreographies less tied to fixed participant sets~\cite{graversen2024alice, samuelson2025quickchanges}; and recent systems explore full-duplex interoperability, distributed dataflow, and restartable actors~\cite{lugovic2024realworld, laddad2024suki, wiersdorf2025chorex}. These systems focus primarily on expressiveness, compilation, and qualitative correctness. \LangSys is complementary: it gives a probabilistic, time-aware semantics for the executions that such systems may generate, including distributions over communication delay, omission, and completion time.

\paragraph{Asynchrony and out-of-order choreographies.}
Several approaches for asynchrony and out-of-order choreographic languages have been proposed. Earlier models relax sequential composition so independent actions can execute without waiting for unrelated earlier actions~\cite{cruzfilipe2017asynchrony, cruzfilipe2018communications}, and recent reasoning principles support verification of choreographic programs~\cite{jongmans2022predicate}. Ozone is the closest prior work to ours: it introduces $O_3$, a model of fully out-of-order choreographies, and an Ozone API that uses futures in Choral to overlap communication and computation while avoiding communication integrity violations~\cite{ozone}. Its semantics is nondeterministic, however, so it explains which reorderings are possible but not how likely they are, how long they take, or how performance depends on communication and computation distributions. \LangSys keeps the same out-of-order intuition but gives every instruction both a value and a timestamp, allowing future-based behavior to be analyzed quantitatively.

\paragraph{Failures and recovery in choreographies.}
Graversen, Montesi, and Peressotti study omission failures in choreographic programming by making communication frames explicit and by giving static robustness analyses for delivery guarantees~\cite{graversen2017promising, graversen2025omission}. Their work is especially relevant to the bounded-retry and two-phase-commit examples in Section~\ref{sec:beyond_koc}: recovery decisions depend on whether messages are observed, and sender and receiver may follow asymmetric recovery strategies. The difference is that their theory treats failure qualitatively and focuses on static guarantees such as at-most-once and exactly-once delivery, whereas \LangSys treats failure, delay, timeouts, and retries as probabilistic effects in the same semantic object. Recent work by Carbone and Veschetti gives a probabilistic choreography language that compiles to PRISM for model checking~\cite{carbone2026probabilistic}. Like \LangSys, this work gives a probabilistic semantics for choreographic programs, but its goal is model checking the \emph{projected} network of processes using PRISM.

\paragraph{Session types.}
Session types and multiparty session types provide a related global view of communication protocols~\cite{honda1998language, honda2016multiparty, huttel2016foundations}. The logical account of sessions connects communication protocols to linear logic~\cite{wadler2012propositions}, and a large body of work uses global and local types to prove communication safety and progress. This differs from choreographic programming in two ways that matter here. First, session types are usually specifications of allowed communication traces rather than executable global programs with dataflow and computation. Second, they typically abstract away from quantitative timing distributions. Probabilistic resource-aware session types~\cite{das2023probabilistic} are particularly close to our goals: they combine session-typed communicating processes with probabilistic choices and static expected-resource reasoning. Their focus is a type system for bounding expected costs of process-level communication, while \LangSys starts from choreographic dataflow and gives the resulting program a probabilistic semantics with explicit timestamps. There is also substantial work on failures and time in session types, including link failures, robust failure handling, exceptional asynchronous sessions, asynchronous message reordering in Rust, and timed multiparty protocols with timeouts~\cite{adameit2017linkfailures,viering2016robust,fowler2019exceptional,cutner2022rumpsteak,hou2024timed}. These systems provide strong static guarantees for typed endpoints. \LangSys instead models a finite program as a probabilistic dependency graph, so a timeout or missing acknowledgement can be used directly as data that controls later instructions.

\paragraph{Process calculi and stochastic process algebras.}
Process calculi such as CSP, CCS, and the $\pi$-calculus provide foundational languages for reasoning about concurrent and mobile communication~\cite{hoare1985csp,milner1989communication,milner1999picalculus}. Asynchronous name-passing calculi make delayed communication explicit and are expressive enough to encode many futures-style or actor-style designs~\cite{merro2004asynchrony}. These calculi are lower-level than choreographies: they describe interacting processes directly, so they do not by themselves provide a global source program with endpoint projection guarantees. Quantitative process algebras such as PEPA add rates and performance models to process-algebraic specifications~\cite{hillston1996pepa}. \LangSys shares the quantitative motivation, but our use of choreographic programming enables interpreting programs using a temporal Bayes net rather than an interleaving process algebra or Markovian transition system. This makes data dependencies, local computations, message delays, omissions, and deadline observations explicit as random variables, and makes reasoning about protocols more tractable.

\paragraph{Temporal Bayesian networks.}
Dynamic Bayesian networks are a mature technique used in machine learning and probabilistic reasoning~\cite{dbn1992}.
Temporal node Bayesian networks are a (slightly) more recent representation, also primarily used in machine learning contexts~\cite{tnbn2005}.
Dataflow programming languages (\eg~\cite{dataflow2004}) are a well-studied domain in their own right,
including non-deterministic versions.
The space of possible use-cases and combinations across these three ideas is wide and rich;
our presentation of a probabilistic time-aware dataflow language,
modeled as a temporal node Bayes net with a proof of correctness of that model,
is novel both in its particulars and in its ability to reason about the performance of
asynchronous choreographies and related distributed computations~\cite{montesi_book}.

\paragraph{Probabilistic programming languages.}
Probabilistic programming languages give programmers general-purpose notation for specifying generative models and running inference~\cite{kozen1981probabilistic, goodman2008church, gordon2014probabilistic}. Systems such as Stan, Pyro, and Gen provide expressive modeling languages and increasingly programmable inference engines~\cite{carpenter2017stan, bingham2019pyro, cusumano2019gen}. \LangSys is deliberately much smaller. It does not aim to be a general statistical modeling language; instead, effect definitions provide the primitive probability distributions, and the program structure composes them into a temporal Bayes net. This restriction is useful for choreographies because it keeps the probabilistic model aligned with communication structure, participant locations, and endpoint execution.

\paragraph{Resource-aware programming languages.}
Resource-aware languages and analyses statically bound program costs such as heap usage, time, or energy. Automatic amortized resource analysis infers symbolic resource bounds through type-based potential annotations, beginning with heap-space bounds for first-order functional programs and extending to multivariate, polynomial, and other resource metrics~\cite{hofmann2003heap,hoffmann2011multivariate,hoffmann2022two}. Ngo et al.~\cite{ngo2018bounded} extend this line to probabilistic programs by deriving symbolic upper bounds on expected resource consumption, and liquid resource types combine refinement typing with amortized resource reasoning to prove precise bounds automatically~\cite{knoth2020liquid}. This work is resource-aware in a different sense: we do not infer worst-case or expected-cost bounds from syntax; instead, \LangSys models resources probabilistically, with time as an explicit random variable. The resulting questions are distributional: expected latency, success probability by a deadline, probability of retransmission, or the distribution of completion times.
Performal~\cite{zhang2023performal} gives a program logic for verifying latency bounds in distributed systems.

\paragraph{Probabilistic model checking with time.}
Timed automata and probabilistic timed automata provide a mature foundation for verification of real-time and stochastic systems~\cite{alur1994timed}. Tools such as PRISM, UPPAAL-SMC, Storm, and Modest support probabilistic or statistical model checking for Markov chains, Markov decision processes, stochastic or priced timed automata, and related models~\cite{kwiatkowska2011prism,bulychev2012uppaal,dehnert2017storm,hensel2022storm,hartmanns2014modest}. These tools are more complete verification environments than \LangSys, and they can check temporal-logic properties against explicit state-space models. \LangSys instead starts from a choreographic program and gives it a probabilistic operational meaning. The two approaches are complementary: a future compiler could translate suitable \LangSys models to a probabilistic model checker, while the direct Bayes-net semantics is convenient for reasoning about program-level latency and failure distributions.

\paragraph{Latency verification for distributed systems.}
 Its examples include systems such as distributed locks, ZooKeeper, and MultiPaxos-based state-machine replication, where the goal is to prove rigorous latency upper bounds for implementations. \LangSys addresses a different level of abstraction: it models choreographic specifications and projected futures-style executions, and it treats latency, omission, retries, and deadline observations probabilistically. Thus Performal is complementary as a verification methodology for deployed systems, while \LangSys provides a probabilistic semantic model for reasoning about the latency distributions induced by choreographic programs.

\section{Conclusion}
\label{sec:conclusion}

We presented \LangSys, a probabilistic, time-aware dataflow language for reasoning about asynchronous choreographies. Its instructions make both values and timestamps explicit via a temporal Bayes-net semantics that gives a quantitative model of the possible executions of a choreography. We showed how this model captures out-of-order choreographic execution and allows reasoning about the execution time of choreographies via an embedding of a subset of $O_3$, and how it can also support settings with inferred knowledge of choice.

\bibliographystyle{plain}
\bibliography{refs}

\begin{acks}
This material is based upon work supported by the Defense Advanced Research Projects Agency (DARPA) under Agreement No. HR00112590082. Any opinions, findings and conclusions or recommendations expressed in this material are those of the author(s) and do not necessarily reflect the views of the United States Government or DARPA.
\end{acks}

\clearpage
\appendix

\input{appendix}
\end{document}

%% file: graphSemantics_large.tex
\section{\LangSys: Asynchronous Instructions Programming}
\label{sec:formal}

We begin by describing the \LangSys language system,
a declarative, probabilistic, time-aware, dataflow language.
In principle, one could implement \LangSys so that its programs can be executed,
but in practice its purpose is to model other systems:
\LangSys programs are interpreted as Bayesian networks, the distribution of which models both the distribution of
values a system may return as well as the time it may take that system to run.
We refer to \LangSys as a "language system"
because most of the semantic meaning of a program is stored in an external environment denoted $\effects$.
\LangSys does not include tools or syntax for constructing an actual $\effects$.
Furthermore, \LangSys is \emph{very simple};
an implementation is free to add whatever macro system they like as a layer between parsing and any of the below analysis.

\subsection{Syntax}
\label{sec:syntax}

Programs in \LangSys are $C$ terms as defined in \Cref{fig:syntax}.
Effectively, a program is just a list of instructions $s$, applied to arguments (which are variables),
and bound to fresh variables.
Every instruction has a guard; its execution is conditional on that guard having been evaluated.
The special variable $\tzero$ has always already been evaluated, so it is natural to write the instruction
$\tzero \ithen y \gets \LangOp{foo}(x)$ as simply $y \gets \LangOp{foo}(x)$;
the examples in Section~\ref{sec:by-example} use this syntactic sugar.
(As a convention, we pre-bind $\tzero: \mathit{Unit} = ((), 0)$).

What effects exist and their type signatures all live outside the program in
an environment $\effects$ of effects, which are indexed by name.
An effect with static parameters (like $\LangOp{com}[p,q]$) can be described as an infinite family of effects,
with the values of their static parameters encoded in their names.

Effects are defined by their names, type signatures, and the probabilistic time-aware functions they represent,
all of which may be written as in prior sections.
For semantic purposes, we access these properties using dot-notation:
An effect $S$ takes some number of arguments $S.\arity$, the types of which are enumerated in $S.\typsIN$,
and it returns a value of type $S.\typOUT$.
$S.\body$ is a function from timestamped (or $\Diamond$) values of the types listed in $S.\typsIN$
to probability distributions over timestamped values of the type given in $S.\typOUT$ (or $\Diamond$).
The boxed effect notation in Section~\ref{sec:by-example} is presentation syntax for such records:
if $s = \LangOp{com}[p,q]$ and $S = \effects[s]$,
then the displayed type signature determines $S.\typsIN$ and $S.\typOUT$,
and the displayed equations define $S.\body$.
Note our use of hat notation ($\WithTimestamp{T}$) does not denote a list of any kind;
$\WithTimestamp{T}$ is shorthand for $(T \times \Timestamps) \cup \Diamond$
(the set of all possible value-timestamp pairs and our null symbol $\Diamond$).
To avoid ambiguity, we use angle braces $\WithTimestamps{\tau s}$ to apply this idea to a list element-wise
instead of treating the whole list as one type.

\begin{mathfigure}{fig:syntax}{The term syntaxes for \LangSys and related structures}
\begin{array}{rcl@{\hspace{4mm}}r}
	x,y,z &\in& \mathbb{V} \triangleq \text{Variables} \cup \{\tzero\}\\
	s &\in& \text{Effect Names}\\
	C &\BNF& I;C
	   \BNFOR 0\\
	I &\BNF& z \ithen y \gets s(x_1, ..., x_N)\\
	\\
	\effects &\in& s \mapsto S\\
	S &\in& \left\{\begin{array}{l|l}
		\mathtt{S} & \mathtt{S}.\arity \in \mathbb{N} \\
		           & \mathtt{S}.\typsIN \in \Types^{\mathtt{S}.\arity}\\
			   & \mathtt{S}.\typOUT \in \Types\\
			   & \mathtt{S}.\body \in \WithTimestamps{\mathtt{S}.\typsIN} \to \Dist(\WithTimestamp{\mathtt{S}.\typOUT})
                \end{array}\right\}\\
	\\
	T &\in& \Types \quad \text{(the set of all types)}\\
	t &\in& \Timestamps \quad \text{(the type of timestamps, which is in $\Types$)}\\
	\Gamma &\in& \mathbb{V} \mapsto \Types\\
	\WithTimestamp{T} &\triangleq& (T\times\Timestamps) \cup \Diamond \\
	\WithTimestamps{T_1, ... T_n} &\triangleq& \WithTimestamp{T_1} \times ... \times \WithTimestamp{T_n}
\end{array}
\end{mathfigure}

\subsection{Type Checking}
\label{sec:typing}

Well-formedness of $C$ terms can only be judged in the context of some $\effects$
that provides the type signatures for the effects.
This is shown in \Cref{fig:typing}.
All there is to check is that the effects' arguments have the correct types and that variables aren't shadowed or overwritten.
By convention, we start from $\Gamma_0 = \varnothing + (\tzero \mapsto \mathit{Unit})$.
In practice when we say $C$ is well-typed, we imply an $\effects$ to be known from context and mean that $\effects, \Gamma_0 \vdash C$.

\begin{mathfigure}{fig:typing}{Typing rules for \LangSys programs.}
\inferrule[TStop]{\,}{\effects, \Gamma \vdash 0}
\\\\
\inferrule[TEffect]
	{
		z \in \Gamma \\
		y \not\in \Gamma\\
		S = \effects[s]\\
		S.\arity = N\\
		\forall 1 \leq n \leq N, \Gamma[x_n] = S.\typsIN[n]\\
		\effects, \Gamma + (y \mapsto S.\typOUT) \vdash C
	}
	{\effects, \Gamma \vdash z \ithen y \gets s(x_1, ..., x_N) ; C}
\\\\
\fbox{Judgments are of the form $\effects, \Gamma \vdash C$}
\end{mathfigure}

\subsection{Centralized Semantics}
\label{sec:semantics}

\begin{mathfigure}{fig:bayes-nets}{Bayes-nets, their nodes, and their well-formedness rules}
\begin{array}{rcl@{\hspace{4mm}}r}
	B &\in& \mathsf{Set}[b] \\
	b &\in& \left\{\begin{array}{l|l}
			\mathtt{b}	& \mathtt{b}.\identifier \in \mathbb{V} \\
					& \mathtt{b}.\typOUT \in \Types\\
					& \mathtt{b}.\arity \in \mathbb{N}\\
					& \mathtt{b}.\varsIN \in \mathbb{V}^{\mathtt{b}.\arity}\\
					& \mathtt{b}.\typsIN \in \Types^{\mathtt{b}.\arity}\\
					& \mathtt{b}.\body \in \WithTimestamps{\mathtt{b}.\typsIN} \to \Dist(\WithTimestamp{\mathtt{b}.\typOUT})
                \end{array}\right\}
		& \text{(nodes)}
\end{array}\\[0.5em]
\fbox{$B + b \triangleq B \cup \{b\}$} \;\;
	\fbox{$B[x] \triangleq b \mathrel{\text{s.t.}} b.\identifier = x \land b \in B$ }\\[0.5em]
\hline\\
\inferrule[BEmpty]{\,}{\cdot\vdash (\varnothing, \varnothing)}
\\\\
\inferrule[BMore]
	{
		\cdot\vdash B \\
		b.\identifier \not\in B \\
		\forall (x, T) \in \mathrm{zip}(b.\varsIN, b.\typsIN), B[x].\typOUT = T
	}
	{\cdot \vdash B + b}
\\\\
\fbox{Judgments are of the form $\cdot\vdash B$}
\end{mathfigure}

\paragraph{Bayesian networks}
\label{sec:bayes-nets}
A Bayesian network is a compact representation of a joint probability distribution.
It consists of a directed acyclic graph whose nodes are random variables.
Each node is equipped with a conditional probability distribution that depends only on the node's parents.
Together, these local conditional distributions determine a joint distribution:
the probability of a complete assignment is the product of the probability of each node's value conditioned on the values of its parents.

This representation is useful here because \LangSys programs are already dependency graphs.
Each instruction binds one variable, and the arguments of the instruction are exactly the values that the bound variable may depend on.
Thus, when we interpret a program as a Bayes net, each program variable becomes a random variable, instruction arguments become parent edges,
and the effect body becomes the conditional distribution for the timestamped value produced by that instruction.
\LangSys represents the behavior of a program as a temporal node Bayes net~\cite{tnbn2005} with uniform discrete time intervals.
\Cref{fig:bayes-nets} shows this representation.
A node $b$ has a variable name as its identifier $b.\identifier$, which is unique and indexable within a (well formed) Bayes net $B$.
A node $b$ additionally has a type $b.\typOUT$,
an ordered list of argument variables $b.\varsIN$,
and a $b.\body$ representing the dependent distribution in the usual Bayes-net sense.
(The arity $b.\arity$ and the types of the arguments $b.\typsIN$ are usually left implicit, as they can be derived from the other attributes.)
A Bayes net $B$ is a DAG over nodes b;
the edge sets are implicit in the argument lists $b.\varsIN$,
and acyclicity is enforced in the well-formedness rule.

\begin{theorem}\label{theorem:well-formed-dags}
	If $\cdot\vdash B$ then $B$ is acyclic and interpretable as a Bayesian Network
	(\ie a joint probability distribution over the values and timestamps represented by its nodes).\\
	This follows directly from the well-formedness rule, which only adds nodes whose parents are already present in the graph.
\end{theorem}

\paragraph{Interpretation of a program as a Bayes Net}
\label{sec:semantic-rules}
Constructing the Bayes net representation $B$ of an \LangSys program $C$ is mostly intuitive.
Every instruction in $C$ will bind one unique variable, which will be represented by a node in $B$.
Rules for this are given in \Cref{fig:semantic-rules}.
Note how the body of an effect $S$ is encapsulated in an off-set transformation
so that it observes time exclusively \emph{after} its guard becomes available.

We define a special node to match our special variable:
$$
\BSemantics{\tzero}{} \triangleq 
\left\{\begin{array}{rcl}
	\identifier &=& \tzero \\
	\typOUT &=& Unit\\
	\varsIN &=& \varnothing\\
	\typsIN &=& \varnothing\\
	\body &=& \singleton((), 0)
\end{array}\right\}
$$
The one root of $B$ will be $\BSemantics{\tzero}{}$,
and the leaves (which one might interpret as "outputs") will correspond to variables that are not used.

\begin{mathfigure}{fig:semantic-rules}{Semantic interpretation of \LangSys programs.}
\begin{array}{rcl}
	(x, t_x) \tplus t &\triangleq& (x, \maximum(t_x + t, 0))\\
	\Diamond \tplus t &\triangleq& \Diamond\\
	\\
	\BSemantics{ 0 }{\effects,B} &\triangleq& B \\
	\BSemantics{  z \ithen y \gets s(x_1, ..., x_N) ;C }{\effects,B} &\triangleq& \BSemantics{ C }{\effects,B + b}\\
	\multicolumn{3}{r}{\text{where } b =
		\left\{ \begin{array}{rcl}
				\identifier &=& y ; \\
				\typOUT &=& \effects[s].\typOUT ; \\
				\typsIN &=& (B[z].\typOUT, ...\effects[s].\typsIN); \\
				\varsIN &=& (z, x_1, ..., x_N) ; \\
				\body(\Diamond, ...) &=& \singleton(\Diamond) \\
				\body((\_, t_z), \WithTimestamp{v_1}, ..., \WithTimestamp{v_N})
				  &=&
				  \hspace{-0.5em}
				  \begin{array}[t]{l l@{\hspace{0.0em}} l}
					  \LangOp{do}
					  & \WithTimestamp{v} \gets \effects[s].\body( & \WithTimestamp{v_1} \tplus (-t_z), \\[-1.0ex]
					  & & ..., \\[-.3ex]
					  & & \WithTimestamp{v_N} \tplus (-t_z)) \\[-0.75ex]
				          & \LangOp{return} \; \WithTimestamp{v} \tplus t_z
				  \end{array}
			\end{array} \right\}
		}
\end{array}\\\\
\fbox{\parbox{0.9\textwidth}{The interpretation of a program is written $\BSemantics{ C }{\effects}$,
	which is shorthand for $\BSemantics{ C }{\effects,\BSemantics{\tzero}{}}$.}}
\end{mathfigure}

The first parent of each Bayes-net node is the instruction's guard.
If the guard is $\Diamond$, the node will also be $\Diamond$.
If the guard resolves at time $t_z$, the rule shifts the effect arguments back by $t_z$ before calling $\effects[s].\body$,
then shifts the result forward by $t_z$.
This makes the effect definitions from Section~\ref{sec:by-example} relative to the time at which the instruction becomes enabled:
an effect that takes one step still takes one step after its guard has arrived.
For example, if a guard becomes available at time $10$ and the effect takes $3$ ticks once enabled,
the effect body computes in a local frame where the instruction starts at time $0$,
and the semantic rule shifts the result back to global timestamp $13$.
Thus effect definitions do not need to know the global time at which they are scheduled.

\begin{theorem}[Soundness of Typing]\label{theorem:type-soundness}
	If $\effects, \Gamma_0 \vdash C$ then $\cdot\vdash \BSemantics{ C }{\effects}$.\\
	(\ie the semantic interpretation of well-typed programs are well-formed Bayes nets.)\\
	For proof, see \Cref{sec:proof-type-soundness}.
\end{theorem}

Bayes nets are just representations of probability distributions.
Sampling from $\BSemantics{ C }{\effects}$ is exactly analogous to
evaluation of $C$ in the "central semantics" of a traditional choreographic language,
and should be thought of as such.
Typically, one will be interested in some sub-set of variables as "outputs".
Marginalizing (by whatever means) $\BSemantics{ C }{\effects}$ to those nodes
is effectively asking "What does this program return?" (and "How long does it take?").

It's likely that some root instructions should be thought of as neither constants nor random variables,
but rather as parameters or inputs.
Parameters should be represented by defining the relevant effects to have no arguments,
constant timestamps,
and any value distribution with sufficiently encompassing support.
Since the distribution chosen is arbitrary, and presumably not accurate, this changes the interpretation of the Bayes net $B$:
One must provide actual values $\mathtt{params}$ for the parameters,
and then ascribe meaning (as above) to the inferred distribution $(B \mid \mathtt{params})$.
Since the inference is strictly forward, this introduces no new challenges.
User inputs may be treated the same way if one is comfortable ascribing them distributions.
Alternately, one might use the Bayes net structure to show
that the marginal distribution one is interested in is independent of the users' inputs.

\subsection{The Network Model}
\label{sec:network}

If all we wanted was to sample from a joint probability distribution,
then much of the above machinery would be unnecessary. 
Our goal is to find distributions that \emph{accurately model} real computations and systems of computations,
and
(in the spirit of choreographic programming)
we would like to avoid duplicating the work of writing our software systems
or leaving the correspondence between theory and practice informal.
That said, modeling the latency of internet communication with any particular distribution will always entail assumptions and approximation.
For example, an assumed distribution of message latencies may become inaccurate when unrelated network traffic changes.

\paragraph{The network world}
\label{sec:real-world}

Operationally, a world state separates the current observations from the work that is still pending.
The store $\Sigma$ records the current timestamped value of each variable, using $\Diamond$ for variables that have not yet become visible.
The map $\Delta$ records pending computations or communications that may still resolve variables.
Each entry in $\Delta$ contains a process $\delta$, which is the computation that may produce a value after its guard is visible.

\begin{mathfigure}{fig:real-world}{A model of real-world computations}
\begin{array}{r@{\hspace{-0.8em}}c@{\hspace{-0.8em}}l@{\hspace{2em}}l}
	W & \BNF \: & (\Delta, \Sigma, t) & \text{(world states)} \\
	\sigma &\in& \bigcup_{T \in \Types} \WithTimestamp{T} \\[1.0ex]
	\delta &\in& \multicolumn{2}{l}{
	  \left\{\begin{array}{l|l}
	      \mathtt{d} & \mathtt{d}.\arity \in \mathbb{N}\\
			 & \mathtt{d}.\typsIN \in (\Types)^{\mathtt{d}.\arity}\\
			 & \mathtt{d}.\typOUT \in \Types\\
			 & \mathtt{d}.\varsIN \in (\mathbb{V})^{\mathtt{d}.\arity}\\
			 & \mathtt{d}.\body \in \WithTimestamps{\mathtt{d}.\typsIN} \times \Timestamps \to \Dist(\mathtt{d}.\typOUT \cup \{\Diamond\})
          \end{array}\right\}
          } \\\\
	\Delta &\in& \mathbb{V} \mapsto \mathbb{V} \times \delta \\
	\Sigma &\in& \mathbb{V} \mapsto \sigma \\
\end{array}\\
\\\hline\\
\Probability{(\Delta, \Sigma, t) \rightsquigarrow (\Delta', \Sigma', t')} \triangleq
\begin{cases}
	t' \neq t + 1 \to & 0 \\
	\exists \Sigma[x] = (v, t) \text{s.t.} \Sigma'[x] \neq (v,t) \to & 0 \\
	\exists \Delta[x] = (z, \delta) \text{s.t.} \Sigma[z] = \Diamond \land \Sigma'[x] \neq \Diamond \to & 0 \\
	\text{otherwise} & p \\
	\exists \Sigma'[x] \neq \Diamond \text{s.t.} x \in \Delta' \to & 0 \\
\end{cases}
\\[1.0ex]
\text{where } p =
	\mathlarger{\mathlarger{\prod}}\limits_{\Delta[x] = (z, \delta)}^{\Sigma[z] = (\_, t_z)}
	\begin{array}{l@{\hspace{0.4em}}l}
	\LangOp{do}
	& d \gets \delta.\body(...\Sigma[\delta.\varsIN] \tplus (- t_z), t - t_z) \\
	& \Probability{(d, t) = \Sigma'[x]}
	  + \Probability{d = \Sigma'[x] = \Diamond}
	\end{array}\\\\
\fbox{Atomic probabilistic time-steps $\rightsquigarrow$ of world states}
\end{mathfigure}

The world, for the purpose of running an \LangSys program,
is a kind of dynamic Bayes net.
Execution begins with a single variable resolved ($\Sigma_0[\tzero] \triangleq ((),0)$),
and a collection $\Delta$ of processes.
For convenience when comparing with the central domain, $\Sigma$ starts containing all variables except $\tzero$ bound to $\Diamond$.
Specifically, $\Delta$ maps each pending variable to its guard variable and its process function $\delta$.
The $\delta$ is not called until the guard is resolved in $\Sigma$.
During each time-step, processes in $\Delta$ whose guards have resolved begin running,
after which they may probabilistically resolve.
A process $\delta$ is \emph{not} an OS process or a $\pi$-calculus process,
it is \emph{any} real-world system that can non-deterministically yield values
from a distribution that depends on some arguments and the current time.
This could be an OS process, or it could be a complicated distributed computation system of its own.
We assume only that these processes can be run in parallel with arbitrary other processes, including copies of themselves, without interference.
A $\delta$ process is less similar to a $S$ effect or a $b$ node than those are to each other;
it represents a within-time process rather than an outside-time distribution.

This representation and the rule for updating from one moment to the next
is shown in \Cref{fig:real-world}.
Semantic stepping is probabilistic.
Some rules about what states a world can step to are encoded as zero-probabilities ascribed to steps to invalid states:
time always increments,
processes in $\Delta$ cannot resolve until their guards resolve,
and once a variable is resolved in $\Sigma$ its process is removed from $\Delta$.
The actual substance of the "steps to" probability distribution is in the "otherwise" case.
Each $\delta$ in $\Delta$ whose guard has resolved is "running", and may resolve or not according to $\delta.\body$.
$\delta.\body$ does not observe universal time, only the time since it started running.
If a process resolves, it is stamped with the time at which it resolved.
(If wall time is desired in a program, one can write an effect to bind a timestamp as data and assign it to a variable.)

We say that a process $\delta$ models a real physical computer system we can run
if that system can read from locations corresponding to each of the $\delta.\varsIN$,
can write to a designated output location (which starts empty and to which the process will write at most once),
and the expected stochastic behavior of the system is correctly described by $\delta.\body$.
In practice, we do not anticipate formal proofs that particular $\delta$ truly model the systems they are claimed to model;
provided a suitable $\delta$ exists, users only need to define the corresponding $S$ effect.

\subsection{Network Projection}
\label{sec:projection}
\label{sec:truncation}

In order to clearly relate the "outside time" perspective of the Bayes Net semantics
to the more realistic time-step semantics, we need a "truncation" operation $\truncate{\cdot}{t}$
that masks off any parts of its argument that aren't available by time $t$.
This is defined in \Cref{fig:truncation}.
Before truncation, $\Diamond$ is a statement about a complete execution: the value never appears.
After truncation, the same symbol can also mean that a value has not appeared yet from the perspective of time $t$.
At its simplest, $\truncate{(v,t_1)}{t_2}$ examines a timestamped value and either masks it to $\Diamond$, or not,
depending on whether it is visible yet;
extension of this operation to apply to $\Diamond$, lists, and dictionaries is straightforward.
For example, a value $(v,10)$ truncates to $\Diamond$ at time $5$, but remains $(v,10)$ once the perspective is after time $10$.
\nb Indexing is perpendicular to truncation; $\truncate{A}{t}[i] = \truncate{A[i]}{t}$;
we use the prior form as often as possible.
In the central Bayes-net semantics, $\Diamond$ means that no timestamped value ever appears.
After truncation, or in the network semantics, the same symbol has an observational meaning:
$\truncate{(v,t_1)}{t_2} = \Diamond$ says only that the value is not visible by time $t_2$.
It may still arrive later.
This distinction is what lets the central semantics talk about complete executions while the network semantics talks about what a process can know at a particular time.

\begin{theorem}{Backward Uniqueness}\label{theorem:backward-uniqueness}
	If $\truncate{a}{t} = \truncate{b}{t}$,
	then for all $t' \leq t$,
	$\truncate{a}{t'} = \truncate{b}{t'}$.

	This follows directly from the definition of truncation.
\end{theorem}

\Cref{fig:truncation} also defines truncation of Bayes nets.
This uses a different notation, $\truncateB{\cdot}{t}$,
because there are two related operations one might mean.
One can sample a complete execution from $B$ and then hide the values that are not visible by time $t$;
this is written $\truncate{B}{t}$ in a $\Probability{\cdot}$ expression.
Alternatively, one can first construct a new Bayes net whose nodes already represent the time-$t$ view of the original execution;
this object is written $\truncateB{B}{t}$.
\Cref{theorem:truncate-bayes} will show that for realizable programs these two distributions coincide.

\begin{mathfigure}{fig:truncation}{Truncation of timestamped data to a moment's perspective.}
\begin{array}{rcl@{\hspace{4mm}}r}
	\truncate{\Diamond}{t} &\triangleq& \Diamond \\
	\truncate{(v,t_1)}{t_2} &\triangleq& \begin{cases} (v, t1) & \text{if $t_1 < t_2$}\\ \Diamond & \text{if $t_1 \geq t_2$} \end{cases} \\
	\truncate{\sigma_1, ..., \sigma_N}{t} &\triangleq& \left( \truncate{\sigma_1}{t}, ..., \truncate{\sigma_N}{t} \right)\\
	\truncate{\left\{ x_1 \mapsto \sigma_1, ..., x_N \mapsto \sigma_N \right\}}{t}
	     &\triangleq& \left\{ x_1 \mapsto \truncate{\sigma_1}{t}, ..., x_N \mapsto \truncate{\sigma_N}{t} \right\} \\
	\hspace{15em}&\hphantom{\triangleq}&\hspace{15em} \\ 
        \hline\\
	\Support(\truncateB{B}{t}) &\triangleq& \left\{ \Sigma' \mid \Sigma' =  \truncate{\Sigma}{t} \land \Sigma \in \Support(B) \right\} \\[0.5em]
	\multicolumn{3}{l}{\Probability{ \truncateB{B}{t}[x] = \truncate{\Sigma}{t}[x]
		\mid \forall_{(y,x) \in \mathrm{edges}(B)} \truncateB{B}{t}[y] = \truncate{\Sigma}{t}[y] } \hspace{10em}} \\
		&\triangleq& \\
	\multicolumn{3}{r}{ \hspace{15em} \Probability{ \truncate{B}{t}[x] = \truncate{\Sigma}{t}[x]
			\mid \forall_{(y,x) \in \mathrm{edges}(B)} B[y] = \Sigma[y] } }
\end{array}
\end{mathfigure}

\Cref{fig:network-semantics} formalizes realizability, the bridge between an outside-time probabilistic effect specification
and an inside-time process.
A central effect $S$ describes a distribution over complete timestamped outputs, while a process $\delta$ runs one time step at a time and sees only what has arrived so far.
The judgment $\delta \Bumpeq S$ says that $\delta$ can generate the same behavior as $S$ without observing future inputs.

\begin{mathfigure}{fig:network-semantics}{Interpretation of an \LangSys program as an initial world state}
\inferrule[REffect]{
	\forall (\sigma^{*} \in \WithTimestamps{T_{ins}},
	         t \in \Timestamps,
		 v \in T_{out}
		),\;
		  \Probability{g(\truncate{\sigma^{*}}{t}, t) = v}
		  = \Probability{f(\sigma^{*}) = (v,t) \mid \truncate{f(\sigma^{*})}{t} = \Diamond}}{
  {\left\{\begin{array}{rcl}
	\typsIN &=& T_{ins} \\
	\typOUT &=& T_{out} \\
	\varsIN &=& \_ \\
	\body &=& g
  \end{array}\right\}}
  \Bumpeq
  {\left\{\begin{array}{rcl}
	\typsIN &=& T_{ins} \\
	\typOUT &=& T_{out} \\
	  \\
	\body &=& f
  \end{array}\right\}}
}
\\\\
\inferrule[RProgram]{
	\effects, \varnothing \vdash C \\
	\forall (z \ithen x \gets s(ys) \in C),\; \Delta[x] \Bumpeq \effects[s]
	\land \Delta[x].\varsIN = ys \\
	\forall (z \ithen x \gets s(ys) \in C),\; ys \cup \{z\} \subset \Support(\Delta) \cup \Support(\Sigma)
	}{
		\effects \vdash (\Delta, \Sigma, 0) \Bumpeq C
	}
\end{mathfigure}

The function attribute $S.\body$ of an effect considers the timestamps of its arguments, but has no knowledge of time itself.
On the other hand, a function $\delta.\body$ can only see previously written values and can only yield a value or not;
it cannot directly set the resulting timestamp.
We bridge these perspectives with judgments $\delta \Bumpeq S$, read \emph{$\delta$ realizes $S$}.
Many of the correspondences in \textsc{REffect} (\Cref{fig:network-semantics}) are presented structurally;
there's also an exact probabilistic relationship between $\delta.\body$ and $S.\body$.

The purpose of this judgment is to rule out central effect definitions that make sense as
Bayes-net nodes but do not describe any implementable distributed behavior.
A central effect $S$ is specified from an outside-time perspective:
its body receives complete timestamped arguments and returns a distribution over complete timestamped results.
Without further restriction, such an effect could use the exact future arrival time of an argument
to decide what value to return now.
A process $\delta$, by contrast, runs from inside time.
At time $t$ it sees only $\truncate{\sigma^{*}}{t}$,
the portion of its arguments that has become visible so far,
and it can either produce a value at the current time or remain unresolved by returning $\Diamond$.
Thus $\delta \Bumpeq S$ says that the outside-time specification $S$
can be generated by some within-time process $\delta$ that never observes future arrivals.

We reuse the $\Bumpeq$ notation for realization of \LangSys programs,
with the added detail that such relation only holds in the context of a $\effects$ to define all the $s$ names.
This restriction is substantive.
The timeout effect from Section~\ref{sec:by-example} is realizable because it waits until deadline $d$ before returning false:
by then, a process can observe whether its argument has arrived.
What realizability rules out is an effect that returns before the deadline while using information about whether the argument will arrive later.
For example, the following central effect is a perfectly well-defined probabilistic function,
but it is not realizable:
\[
\boxed{
\begin{aligned}
&\LangOp{late}[d] &:&\quad (\tau \times \Timestamps \cup \Diamond \to \Dist(\mathit{Bool} \times \Timestamps \\
&\LangOp{late}[d](\Diamond) &=&\quad \textbf{return}\; (\LangOp{true}, 0)\\
&\LangOp{late}[d]((v,t)) &=&\quad \textbf{return}\; \begin{cases}
	(\LangOp{true}, 0) & \text{if } t > d \\
	(\LangOp{false}, 0) & \text{otherwise}
\end{cases}
\end{aligned}
}
\]
This effect returns at time $0$ while deciding whether its argument will arrive after deadline $d$
(or never arrive at all).
No distributed process can implement this behavior:
before the deadline, all not-yet-arrived inputs have the same truncated view, namely $\Diamond$,
so a within-time process cannot distinguish an input that will arrive late from one that will arrive soon.
Realizability is the condition that precludes this kind of time travel.
The next theorem states the corresponding no-foresight property:
for realizable programs, what is visible by time $t$ depends only on what the inputs look like by time $t$,
not on their future values or arrival times.

\begin{theorem}\label{theorem:no-foresight}
	If $C$ is realizable under $\effects$,
	\ie there exists some $\Delta$ s.t. $\effects \vdash W \Bumpeq C$,
	then for all $z \ithen x \gets s(ys) \in C$,
	for all $t$ and $\sigma$,
	and for all $\Sigma \in \Support(\BSemantics{C}{\effects}[ys])$:
	$$
	\Probability{ \truncate{\BSemantics{C}{\effects}}{t}[x] = \sigma \mid \BSemantics{C}{\effects}[ys,z] = \Sigma }
	=
	\Probability{ \truncate{\BSemantics{C}{\effects}}{t}[x] = \sigma \mid \truncate{\BSemantics{C}{\effects}}{t}[ys,z] = \truncate{\Sigma}{t} }
	$$
	Proof is presented in \Cref{sec:proof-no-foresight}.
	Note that this property can be used transitively for a $\Sigma'$ if $\truncate{\Sigma'}{t} = \truncate{\Sigma}{t}$.
\end{theorem}

We can now address the subject of a truncated Bayes net,
as defined in \Cref{fig:truncation}.
The joint probability distribution of a truncated Bayes net is defined piecewise for each node (variable) conditional on its parents,
therefore $\truncateB{B}{t}$ is in general a well-defined Bayes net
(at least if $B = \BSemantics{C}{\effects}$ for some $C$).
We proceed to show that if that $C$ is realizable, then sampling from the truncated Bayes net is the same as truncating samples
from the original Bayes net;
the distributions are the same.

\begin{theorem}[Bayes net truncation is sound.]\label{theorem:truncate-bayes}
If $B = \BSemantics{C}{\effects}$ and there exists a $W$ s.t. $\effects \vdash W \Bumpeq C$, then
$$
\Probability{ \truncateB{B}{t} = \Sigma }
=
\Probability{ \truncate{B}{t} = \Sigma }
$$
We prove this in \Cref{sec:proof-truncate-bayes}.
\end{theorem}

The purpose of all of this machinery is to show that the Bayes-net interpretation of a program captures
its distribution of possible behaviors.

\begin{theorem}[Soundness of realizations]\label{theorem:real-soundness}
	If $\effects, \Gamma_0 \vdash C$,
	and $\effects \vdash W_0 \Bumpeq C$,
	and $\Probability{W_0 \rightsquigarrow^* (\Delta_1, \Pi_1 \Sigma_1, t_1)} \neq 0$,
	then
	$$
	\Probability{ (\Delta_1, \Sigma_1, t) \rightsquigarrow (\Delta_2, \Sigma_2, t+1) }
	=
	\Probability{ \truncate{\BSemantics{C}{\effects}}{t+1} = \Sigma_2 \mid \truncate{\BSemantics{C}{\effects}}{t} = \Sigma_1}
	$$
	The right-hand side is the product of the probabilities of all the variables resolving to $(v,t)$
	(or something that truncates to $\Diamond$) given that they do not resolve at any $t'<t$.
	\textsc{REffect} ensures that the relative distribution is captured by the corresponding $\delta$.
	The probabilistic time-step definition replicates the offsetting of each $S.\body$ by the guard variables,
	and products together the correct set of variables.
\end{theorem}

Our expectation is that in practice,
users will skip directly from real software they wish to model to $S$ effects that they are comfortable with as models.
As long as a suitable $\delta$ \emph{exists}, it is not necessary to construct it.

%% file: appendix.tex
\section{Proofs}
\label{sec:proofs}

\subsection{Well-typed programs have Bayes-net semantics.}
\label{sec:proof-type-soundness}
Recall \Cref{theorem:type-soundness}, which says that
if $\effects, \Gamma_0 \vdash C$ then $\cdot\vdash \BSemantics{ C }{\effects, \BSemantics{\tzero}{}}$.
We will need a notion of agreement between $\Gamma$ contexts and (preexisting) Bayes nets:

\begin{align*}
\Gamma \risingdotseq B \quad \triangleq 
	\hphantom{\land} & \forall x \mapsto T  \in \Gamma, B[x].\typOUT = T \\
	\land            & \forall b \in B, b.\identifier \in \Gamma
\end{align*}

This lets us generalize our theorem as:
If $\effects, \Gamma \vdash C$ and $\cdot\vdash B$ and $\Gamma \risingdotseq B$ then $\cdot\vdash \BSemantics{ C }{\effects, B}$.
(Since $\Gamma_0 \risingdotseq \BSemantics{\tzero}{}$, the original theorem is a special case of this.)
The empty program is well typed for any $\Gamma$, and its interpretation does not update the existing graph, so the base case is immediate.
To complete the inductive proof we consider some $I;C$ and enumerate some assumptions:
\begin{enumerate} \setcounter{enumi}{-1}
	\item $I = z \ithen y \gets s(x_1, ..., x_{\arity})$
	\item $\cdot\vdash B$
		(We are proving a conditional, so we assume its prerequisites, but not in order.)
	\item $\Gamma \risingdotseq B$
		(same)
	\item $\varsIN = [x_1, ..., x_{\arity}]$ (Notational convenience. We will implicitly cast this to a set as needed.)
	\item $S = \effects[s] =
		\left\{  \;
		\arity   \;;\;
		\typsIN   \;;\;
		\typOUT    \;;\;
		\body    \;;\;
		\tims    \;
		\right\}
		$
	\item $\Gamma' = \Gamma \oplus\{y \mapsto \typOUT\}$
	\item (The conditional's third prerequisite)
		The program is typed by \textsf{TEffect} (and so that rule's prerequisites are also taken by assumption):
$$
\inferrule
	{
		z \in \Gamma\\
		y \not\in \Gamma\\
		\Gamma[\varsIN] = \typsIN\\
		\effects, \Gamma' \vdash C
	}
	{\effects, \Gamma \vdash I ; C}
$$
	\item Our inductive hypothesis: for all $\Gamma_z$ and $B_z$,
		if $\effects, \Gamma_z \vdash C$ and $\cdot\vdash B_z$ and $\Gamma_z \risingdotseq B_z$
		then $\cdot\vdash \BSemantics{ C }{\effects, B_z}$.
\end{enumerate}

What we want to show is that 
$
\cdot\vdash \BSemantics{ C }{\effects, B + b}
$
where $b$ is constructed according to the semantic definition in \Cref{fig:semantic-rules}.

We get $\effects, \Gamma' \vdash C$ from \#6.

To show $\cdot\vdash B'$ requires applying \textsc{BMore}.
We already have $\cdot\vdash B$ (\#1).
$y \not\in \Gamma$ (\#6) implies the respective absence in $B$, because $\Gamma \risingdotseq B$ (\#2).
The argument names $ids$ are constructed to be correct and similarly present and typed in $B$.

Since $\Gamma \risingdotseq B$ (\#2) and $\Gamma'$ and $B'$ are each just extended by $y$,
simple inspection shows that $\Gamma' \risingdotseq B'$.

The above combine via \#7 to complete the inductive case.

\subsection{Realizable effects cannot see the future}\label{sec:proof-no-foresight}
Recall our target:
If $C$ is realizable, then conditioning the truncation of some $x$ on exact values of its parents is the same as
conditioning on the truncation of those same parent values.

Suppose the truncated $x$ value whose probability we are checking is some $(v,t')$ where $t'<t$.
Then the probability of observing that $(v,t')$ conditional on exact parents $\sigma^{*}$ of $x$ is taken directly $\BSemantics{C}{\effects}[x].\body$.
Since there is some $\delta$ which realizes the corresponding effect,
the probability must be the same for any $\sigma'$ s.t. $\truncate{\sigma'}{t'} = \truncate{\sigma^{*}}{t'}$,
a superset of the possible values that agree up to $t$.
Since the probability is the same for any $\sigma'$ s.t. $\truncate{\sigma'}{t} = \truncate{\sigma^{*}}{t}$,
conditioning on the weaker condition is the same as conditioning on exact values.

If we suppose instead that the truncated $x$ value whose probability we are checking is $\Diamond$,
then the probability is simply the complement of the sum of the probabilities of all $(v,t')$ where $t'<t$,
so no further proof is necessary.

\subsection{Sampling from a truncated Bayes net is the same as truncating the samples.}\label{sec:proof-truncate-bayes}
\Cref{theorem:truncate-bayes} assumes that $B$ is the Bayes-net interpretation of a realizable program.
The support of a truncated Bayes net cannot contain timestamps past its "present" time,
which we can denote by representing the value in question as a truncation of some unspecified $\Sigma$.
It suffices to show that:
$$
\Probability{ \truncateB{B}{t} = \truncate{\Sigma}{t} }
=
\Probability{ \truncate{B}{t} = \truncate{\Sigma}{t} }
$$

We begin with the factorization definition of a bayes net, since we know the truncation is one:
$$
\Probability{ \truncateB{B}{t} = \truncate{\Sigma}{t} }
=
\product_{ys \mapsto x \in B}
\Probability{ \truncateB{B}{t}[x] = \truncate{\Sigma}{t}[x]
	\mid \truncateB{B}{t}[ys] = \truncate{\Sigma}{t}[ys] }
$$
$$
\triangleq
\product_{ys \mapsto x \in B}
\Probability{ \truncate{B}{t}[x] = \truncate{\Sigma}{t}[x]
	\mid B[ys] = \Sigma[ys] }
$$
$$
=
\product_{ys \mapsto x \in B}
\Probability{ \truncate{B}{t}[x] = \truncate{\Sigma}{t}[x]
	\mid \truncate{B}{t}[ys] = \truncate{\Sigma}{t}[ys] }
\quad\text{\Cref{theorem:no-foresight}}
$$

To show that this is equal to $\Probability{ \truncate{B}{t} = \truncate{\Sigma}{t} }$,
we do induction on the structure of $B$.
\begin{itemize}
	\item \textbf{Base case:}
		For all $z \ithen x \gets s(ys) \in C$, $ys = \varnothing$ and $z = \tzero$.\\
		Since all of the variables are independent
		(and there are no conditionals being applied),
		we can simply use the multiplication rule for independent variables
		to convert the product of probabilities to a probability of conjunction,
		which rewrites as our target.
	\item \textbf{Inductive case:}
		$B = B_0 + \{b\}$
		and $b.\varsIN \subseteq B_0$ and $b.\identifier = x_1$.
		Our inductive hypothesis is that
		$$
		\product_{ys_0 \mapsto x_0 \in B_0}
		  \Probability{ \truncate{B_0}{t}[x_0] = \truncate{\Sigma}{t}[x_0]
		  \mid \truncate{B_0}{t}[ys_0] = \truncate{\Sigma}{t}[ys_0] }
		=
		\Probability{ \conjunction_{x_0 \in B_0} \truncate{B_0}{t}[x_0] = \truncate{\Sigma}{t}[x_0] }
		$$
		From there we can pick up with
		$$
		\begin{aligned}
			&  \product_{ys \mapsto x \in B}
				\Probability{ \truncate{B}{t}[x] = \truncate{\Sigma}{t}[x]
					\mid \truncate{B}{t}[ys] = \truncate{\Sigma}{t}[ys] } \\[.5em]
			=& \Probability{ \truncate{B}{t}[x_1] = \truncate{\Sigma}{t}[x_1]
					\mid \truncate{B}{t}[ys_1] = \truncate{\Sigma}{t}[ys_1] } \\
			 & \times \product_{ys_0 \mapsto x_0 \in B_0}
				\Probability{ \truncate{B}{t}[x_0] = \truncate{\Sigma}{t}[x_0]
					\mid \truncate{B}{t}[ys_0] = \truncate{\Sigma}{t}[ys_0] } \\[.5em]
			\text{inductive h.}\quad
			=& \Probability{ \truncate{B}{t}[x_1] = \truncate{\Sigma}{t}[x_1]
					\mid \truncate{B}{t}[ys_1] = \truncate{\Sigma}{t}[ys_1] } \\
			 & \times \Probability{ \conjunction_{x_0 \in B_0} \truncate{B_0}{t}[x_0] = \truncate{\Sigma}{t}[x_0] } \\[.5em]
			\text{total pr.\&cond.}\quad
			=& \left( \sumation_{\truncate{\Sigma_{ys1}}{t} = \truncate{\Sigma}{t}[ys_1]} \begin{aligned}
				\hphantom{\times} &
				\Probability{ \truncate{B}{t}[x_1] = \truncate{\Sigma}{t}[x_1]
					\mid B[ys_1] = \Sigma_{ys1} } \\
				\times & \Probability{ B[ys_1] = \Sigma_{ys1} \mid \truncate{B}{t}[ys_1] = \truncate{\Sigma}{t}[ys_1] }
				\end{aligned}\right)\\
			 & \times \Probability{ \conjunction_{x_0 \in B_0} \truncate{B_0}{t}[x_0] = \truncate{\Sigma}{t}[x_0] }
		\end{aligned}
		$$
		Let $\Sigma_{0s}$ be the \emph{set} of sigma-dictionaries defined over the variables of $B_0$ s.t.
		for all $x_0 \in B_0$ and $\Sigma_0 \in \Sigma_{0s}$, $\truncate{\Sigma_0}{t}[x_0] = \truncate{\Sigma}{t}[x_0]$.
		$$
		\begin{aligned}
			=& \left( \sumation_{\truncate{\Sigma_{ys1}}{t} = \truncate{\Sigma}{t}[ys_1]} \begin{aligned}
				\hphantom{\times} &
				\Probability{ \truncate{B}{t}[x_1] = \truncate{\Sigma}{t}[x_1]
					\mid B[ys_1] = \Sigma_{ys1} } \\
				\times & \Probability{ B[ys_1] = \Sigma_{ys1} \mid \truncate{B}{t}[ys_1] = \truncate{\Sigma}{t}[ys_1] }
				\end{aligned}\right) \\
			 & \times \left( \sumation_{\Sigma_0 \in \Sigma_{0s}}
				\Probability{ \conjunction_{x_0 \in B_0} B_0[x_0] = \Sigma_0[x_0] }
				\right)
		\end{aligned}
		$$
		Since $B_0$ is just a sub-DAG of $B$, we can replace replace a reference to it above.
		That and some re-arranging gets us
		$$
		\begin{aligned}
			=& \sumation_{\truncate{\Sigma_{ys1}}{t} = \truncate{\Sigma}{t}[ys_1]}
			   \sumation_{\Sigma_0 \in \Sigma_{0s}}
			\left(\begin{aligned}
				\hphantom{\times} &
				\Probability{ \truncate{B}{t}[x_1] = \truncate{\Sigma}{t}[x_1]
					\mid B[ys_1] = \Sigma_{ys1} } \\
				\times & \Probability{ B[ys_1] = \Sigma_{ys1} \mid \truncate{B}{t}[ys_1] = \truncate{\Sigma}{t}[ys_1] } \\
				\times & \Probability{ \conjunction_{x_0 \in B_0} B[x_0] = \Sigma_0[x_0] }
			\end{aligned}\right)
		\end{aligned}
		$$
		$\Sigma_{ys1}$ and $\Sigma_0$ are both defined to agree with $\Sigma$ over $ys_1$ under truncation to $t$,
		so we can use \Cref{theorem:no-foresight}:
		$$
		\begin{aligned}
			=& \sumation_{\truncate{\Sigma_{ys1}}{t} = \truncate{\Sigma}{t}[ys_1]}
			   \sumation_{\Sigma_0 \in \Sigma_{0s}}
			\left(\begin{aligned}
				\hphantom{\times} &
				\Probability{ \truncate{B}{t}[x_1] = \truncate{\Sigma}{t}[x_1]
					\mid B[ys_1] = \Sigma_0[ys_1] } \\
				\times & \Probability{ B[ys_1] = \Sigma_{ys1} \mid \truncate{B}{t}[ys_1] = \truncate{\Sigma}{t}[ys_1] } \\
				\times & \Probability{ \conjunction_{x_0 \in B_0} B[x_0] = \Sigma_0[x_0] }
			\end{aligned}\right)
		\end{aligned}
		$$
		Since we are conditioning $x_1$ on exact values of its parents,
		and all of $ys_1$ are in $B_0$ (and no descendants are),
		we can combine the first and third terms:
		$$
		\begin{aligned}
			=& \sumation_{\truncate{\Sigma_{ys1}}{t} = \truncate{\Sigma}{t}[ys_1]}
			   \sumation_{\Sigma_0 \in \Sigma_{0s}}
			\left(\begin{aligned}
				\hphantom{\times} &
				\Probability{ \truncate{B}{t}[x_1] = \truncate{\Sigma}{t}[x_1]
					\land  \conjunction_{x_0 \in B_0} B[x_0] = \Sigma_0[x_0] } \\
				\times & \Probability{ B[ys_1] = \Sigma_{ys1} \mid \truncate{B}{t}[ys_1] = \truncate{\Sigma}{t}[ys_1] }
			\end{aligned}\right)
		\end{aligned}
		$$
		Factor out the later term and convert the inner summation back into a disjunction,
		\ie equality under truncation:
		$$
		\begin{aligned}
			=& \sumation_{\truncate{\Sigma_{ys1}}{t} = \truncate{\Sigma}{t}[ys_1]}
			\left(\begin{aligned}
				\hphantom{\times} &
				\Probability{ B[ys_1] = \Sigma_{ys1} \mid \truncate{B}{t}[ys_1] = \truncate{\Sigma}{t}[ys_1] } \\
				\times &
				\Probability{ \truncate{B}{t}[x_1] = \truncate{\Sigma}{t}[x_1]
					\land \conjunction_{x_0 \in B_0} \truncate{B}{t}[x_0] = \truncate{\Sigma}{t}[x_0] }
			\end{aligned}\right)
		\end{aligned}
		$$
		$x_1$ is the only variable missing from $B_0$ in $B$,
		so the conjunction can be rewritten and factored out.
		Without it, the summation is just $1$, and the result follows:
		$$
		= \Probability{ \truncate{B}{t} = \truncate{\Sigma}{t} }
		$$
\end{itemize}